%% file: main.tex
\documentclass[superscriptaddress, twocolumn]{revtex4-2}
\usepackage[english]{babel}
\usepackage[utf8]{inputenc}
\input{preamble}

\usepackage[pdftex, pdftitle={Article}, pdfauthor={Author}]{hyperref} 
\usepackage[capitalize,nameinlink]{cleveref}
\hypersetup{
	colorlinks=true, 
	linkcolor=black,  
	citecolor=blue, 
}

\begin{document}
	\title{Simulating near-infrared spectroscopy on a quantum computer for enhanced chemical detection}
	
	\author{Ignacio Loaiza}
	\thanks{These authors contributed equally. \\
		stepan.fomichev@xanadu.ai\\
		ignacio.loaiza@xanadu.ai}
	\affiliation{Xanadu. Toronto, ON. M5G 2C8. Canada}
	\author{Stepan Fomichev}
	\thanks{These authors contributed equally. \\
		stepan.fomichev@xanadu.ai\\
		ignacio.loaiza@xanadu.ai}
	\affiliation{Xanadu. Toronto, ON. M5G 2C8. Canada}
    \author{Danial Motlagh}
	\affiliation{Xanadu. Toronto, ON. M5G 2C8. Canada}
    \author{Pablo A. M. Casares}
	\affiliation{Xanadu. Toronto, ON. M5G 2C8. Canada}
    \author{Daniel Honciuc Menendez}
    \affiliation{Xanadu. Toronto, ON. M5G 2C8. Canada}
	\affiliation{Department of Physics, University of Toronto. Toronto, ON. M5S 1A1. Canada}
     \author{Serene Shum}
     \affiliation{Xanadu. Toronto, ON. M5G 2C8. Canada}
	\affiliation{Department of Physics, University of Toronto. Toronto, ON. M5S 1A1. Canada}
	\author{Alain Delgado}
    \affiliation{Xanadu. Toronto, ON. M5G 2C8. Canada}
	\author{Juan Miguel Arrazola}
	\affiliation{Xanadu. Toronto, ON. M5G 2C8. Canada}
	
	\newcommand{\be}{\mathbb{E}}
	
	\date{\today}
	
	\begin{abstract}
		Near-infrared (NIR) spectroscopy is a non-invasive, low-cost, reagent-less, and rapid technique to measure chemical concentrations in a wide variety of sample types. However, extracting concentration information from the NIR spectrum requires training a statistical model on a large collection of measurements, which can be impractical, expensive, or dangerous. In this work, we propose a method for simulating NIR spectra on a quantum computer, as part of a larger workflow to improve NIR-based chemical detection. The quantum algorithm is highly optimized, exhibiting a cost reduction of many orders of magnitude relative to prior approaches. The main optimizations include the localization of vibrational modes, an efficient real-space-based representation of the Hamiltonian with a quantum arithmetic-based implementation of the time-evolution, optimal Trotter step size determination, and specific targeting of the NIR region. Overall, our algorithm achieves a $\mathcal{O}(M^2)$ scaling, compared with the $\mathcal{O}(M^{12})$ coming from equivalent high-accuracy classical methods. As a concrete application, we show that simulating the spectrum of azidoacetylene (HC$_2$N$_3$), a highly explosive molecule with strong anharmonicities consisting of $M=12$ vibrational modes, requires circuits with a maximum $8.47\times10^8$ T gates and $173$ logical qubits. By enhancing the training datasets of detection models, the full potential of vibrational spectroscopy for chemical detection could be unlocked across a range of applications, including pharmaceuticals, agriculture, environmental monitoring, and medical sensing.
	\end{abstract}

	\maketitle
		
	\glsresetall
	\section{Introduction}

Detection of harmful chemicals in the air, water and soil is a crucial first step to tackling environmental and industrial pollution \cite{slonecker2010visible,wilson2012review,tan2019scheme,thakur2022recent}. Near-infrared absorption spectroscopy (NIRS) is particularly well-suited to this task \cite{huck2006near,sulub2007spectral,siesler2008near,ozaki2018near,bik2020lipid,bec2022silico}: it is a non-destructive, ultrafast, reagent-less, easily miniaturized technique with broad applicability across different media, with no sample preparation requirements. Wide adoption of NIRS for tracking harmful chemicals could deliver real-time spatiotemporal data essential for addressing challenges as diverse as disaster relief \cite{jha2008advances}, disease diagnosis \cite{adegoke2021near}, agricultural soil quality monitoring \cite{cecillon2009assessment}, medical sensing \cite{medical}, and climate change \cite{counsell2016recent,corradini2019predicting,zhao2023role}. 

While the near-infreared (NIR) spectrum is rich in chemical-identifying information, this information is nontrivial to extract \cite{sulub2007spectral,palafox2017computational,bec2019breakthrough,ozaki2021near}. The NIR region of $12,500-4,000$ cm$^{-1}$ ($800-2,500$ nm) is dominated by combination and overtone vibrational bands. These bands often include large anharmonic effects that make band assignment from harmonic frequencies virtually impossible. In addition, the combinatorial number of bands yields broad distributions of intensities across many wavelengths. \textit{Spectral fingerprinting} in this NIR region thus requires to go beyond peak identification and incorporate information from the entire spectrum. It is thus necessary to collect a large amount of spectral and concentration data, and then use it to train statistical models to correlate certain global spectral patterns with chemical concentration \cite{beebe1998chemometrics,ozaki2021near}. Such data collection is often time-consuming \cite{nagy2022quality}, expensive \cite{grabska2017temperature,sibert2023modeling}, and even dangerous \cite{tan2019scheme,van2023rapid}, depending on the chemical of interest.

Simulating NIR spectra is a promising way to augment NIR training datasets \cite{westad2008incorporating,bec2018nir,bec2019simulated,ozaki2021advances}. It is inherently safe, inexpensive, and has the potential to alleviate many of the challenges associated with collecting training data. However, while fundamental frequencies appearing in the \textit{mid}-IR region ($4,000-400$ cm$^{-1}$) can be simulated with sufficient accuracy by inexpensive methods like harmonic analysis \cite{bec2021introduction}, the presence of combination and overtone bands in NIR necessitates the use of much more computationally demanding techniques such as vibrational configuration interaction (VCI) \cite{ozaki2021advances,oschetzki2013azidoacetylene}. While VCI and similar approaches are capable of accurately predicting peak positions and intensities of such bands, the associated runtime and memory cost scales prohibitively with system size, motivating the need for alternative approaches. This steep cost scaling has thus far restricted the applicability of NIR simulations to industrially useful contexts \cite{bec2021introduction}.

\begin{figure*}[t]
\centering
\includegraphics[width=\linewidth]{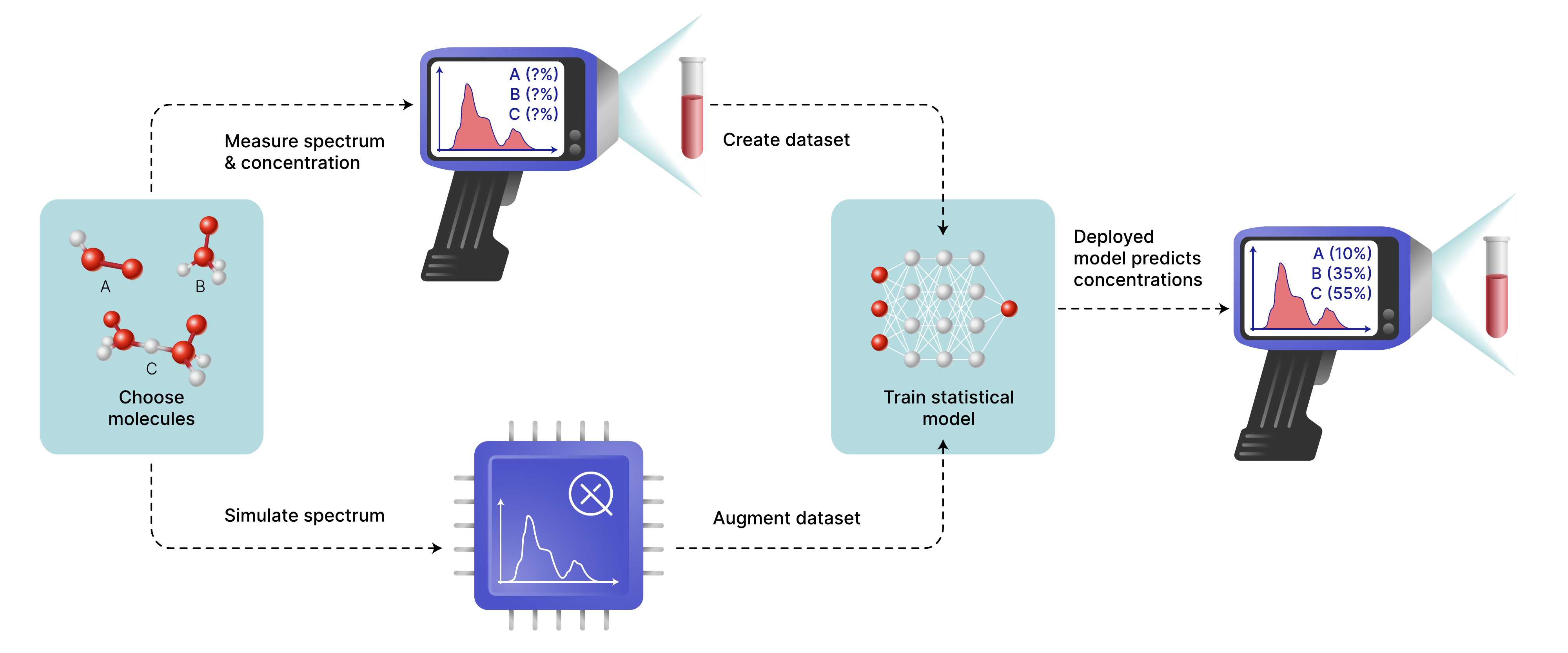}
\caption{\textbf{Chemical detection using near-infrared absorption spectra}. Once a molecule is chosen, one typically needs to collect many measurements of varying concentrations and under different conditions in order to train a statistical model that can then infer the concentration of the molecule by detecting a signature of its spectrum in an unknown sample. We propose augmenting the training dataset and training process of the statistical model using simulated NIR spectra obtained on a quantum computer. }
\label{fig:workflow}
\end{figure*}

In contrast to the steep scaling from high-accuracy classical solutions, the usage of quantum computers presents a promising avenue for escalating calculations to larger systems. However, previous estimates for the cost of quantum-based simulation of vibrational systems were beyond the reach of early fault-tolerant hardware \cite{ibm_vibrational}, motivating the need for more efficient quantum algorithms. In this work, we propose a novel quantum algorithm for simulating vibrational spectra of molecules. The algorithm is shown to deliver high-accuracy NIR spectra at significantly more affordable computational cost and favourable scaling compared to classical approaches of similar accuracy. The algorithm is highly optimized to decrease constant factors in the runtime, achieving a reduction of many orders of magnitude compared with earlier quantum methods \cite{ibm_vibrational}, and already becoming competitive in runtime with state-of-the-art \textit{classical} techniques for modestly-sized molecules with only seven to ten atoms. 

The optimizations used in the algorithm consist of the following: mode localization techniques for reducing the required number of terms in the Hamiltonian while achieving high-accuracy \cite{mode_loc_1,mode_loc_2,mode_loc_3,mode_loc_4}, a real-space representation of the vibrational Hamiltonian \cite{macridin_1,macridin_2}, alongside its associated quantum-arithmetic-based implementation of the Trotterized time-evolution operator (adapted from the vibronic algorithm in Ref.~\cite{vibronic}), a novel and accurate estimation of the Trotter step size based on the perturbative approach from Refs.~\cite{trotter_pt, trotter_pt_2}, a time-domain quantum algorithm \cite{qpe_gqpe} with a few compilation-based optimizations to reduce its overall cost \cite{new_xas}, an optimal window selection plus initial state preparation targeting the NIR region, and usage of the active volume compilation technique \cite{active_volume}. The overall improvements from these optimizations are summarized in \cref{tab:improvements} for the example of the azidoacetylene HC$_2$N$_3$ molecule.
\begin{table*}[]
    \centering
    \begin{tabular}{|l|c|c|c|c|}
       \hline  \textbf{Optimization technique} & \textbf{Additional qubits} & \textbf{Total qubits}& \begin{tabular}{@{}c@{}}\textbf{Approximate} \\ \textbf{improvement}\end{tabular}&  \begin{tabular}{@{}c@{}}\textbf{Cumulative} \\ \textbf{improvement}\end{tabular}\\ \hline \hline
         1 - Mode localization & $0$ & $49$ & $\times 50$& $5\times 10^1$\\ \hline
         2 - Real-space representation & $124$& $173$& $\times 8$& $4\times 10^2$\\ \hline
     3 - Coefficient caching & $0$& $173$& $\times 2$ & $8\times 10^2$\\ \hline
     4 - Trotter perturbative step & $0$ & $173$& $\times 2$ & $1.6\times 10^3$\\ \hline
     5 - Double measurement trick & $0$& $173$& $\times 1.5$& $2.4\times 10^3$\\ \hline
     6 - Double phase trick& $0$ & $173$&  $\times 2$ & $4.8\times 10^3$\\ \hline
     7 - Initial state projection& $0$ & $173$& $\times 1000$& $4.8\times 10^6$\\ \hline
     8 - Active volume compilation & $327-1827$& $500-2000$ & $\times 25-\times100$ & $1.2\times 10^8-4.8\times10^8$\\ \hline
     \end{tabular}
    \caption{List of optimizations with associated estimated improvement ratio to the total runtime required to simulate NIR spectra. The cumulative improvement column corresponds to an aggregate of all optimizations on top of the associated row, while total qubits corresponds to the associated number of qubits. Numbers shown are obtained for the azidoacetylene HC$_2$N$_3$ molecule with $12$ vibrational modes; some optimizations will have a larger improvement for bigger molecules. The mode localization improvement was calculated through the change of number of terms in the Hamiltonian obtained with a $2$-mode expansion instead of $3$-mode. The real-space representation improvement was calculated with respect to the second-quantized modal-based representation of the Hamiltonian \cite{christiansen,ibm_vibrational}. The Trotter step improvement was calculated with respect to the usage of the commutator-based bounds \cite{trotter_error}. Two different budgets of 500 and 2000 logical qubits were considered with the active volume compilation technique to showcase the different improvements.}
    \label{tab:improvements}
\end{table*}

This paper is organized as follows. In \cref{sec:application}, we describe how NIRS is used in practice for chemical detection, highlight the potential benefit of simulations, and elaborate on the challenges faced by existing simulation methods, classical and quantum. We then present the new algorithm in \cref{sec:algorithm}, where we lay out all the optimizations that allow us to achieve the large runtime reduction. The algorithm is then benchmarked through proof-of-concept simulations in \cref{sec:benchmark} -- simulations which also allow us to perform robust and careful constant-factor resource estimation. Our conclusions are presented in \cref{sec:conclusions}.

\section{Near-infrared spectroscopy for chemical detection}

\label{sec:application}

\subsection{Practical spectral fingerprinting with NIRS}
When NIR radiation is passed through a sample, it is partially absorbed by the constituent molecules. NIR radiation is energetically too weak to generate electronic excitations, as happens for example with UV-visible or X-ray spectroscopy: instead, incoming radiation is translated into vibrations of the atoms in the absorbing molecule. The key quantity measured in NIRS experiments is the absorption cross-section $\sigma_A(\omega)$, a function indicating the strength of absorption at a particular frequency of incoming radiation $\omega$. Within the dipole approximation, it is given by    
\begin{equation} \label{eq:absorption}
    \sigma_A(\omega) \propto \omega \sum_{\rho=x,y,z}\sum_{f} |\bra{E_f} \hat\mu_\rho \ket{E_0}|^2 \mathcal{L}_\eta(\omega-E_f+E_0).
\end{equation}
Here $\ket{E_0}, \ket{E_f}$ are the ground and excited states of a molecular vibrational Hamiltonian $\hat{H}$, $\omega$ is the frequency of incoming radiation, $\hat{\mu}_{\rho}$ is the $\rho$-th Cartesian component of the molecular dipole operator, and $\mathcal{L}_{\eta}(\omega) = \eta / (\omega^2 + \eta^2)$ is the Lorentzian broadening with $\eta$ its half-width at half-maximum. This broadening is related to the lifetime of excited states of the system as well as the resolution of the measurement apparatus, and in practice usually takes values between $\eta=1$ cm$^{-1}$ and $10$ cm$^{-1}$.

The set of peak positions $E_f$ and absorption intensities $|\bra{E_f} \hat{\mu}_{\rho} \ket{E_0}|^2$ is unique to a given molecule, being determined by the details of its ground-state potential energy surface (PES). Since peaks and their intensities completely determine the absorption spectrum, that implies the spectrum itself is a unique \textit{fingerprint} of the molecule. Thus if molecular spectra are cataloged \textit{a priori}, it should be possible to detect the presence of a given molecule in an unknown sample through a direct comparison against the catalog. This is indeed how chemical detection occurs with mid-IR spectra. The peaks present in the mid-IR range ($4,000-400$ cm$^{-1}$) are fundamental vibrational excitations: there is usually no shortage of sharp, high-intensity features that are highly specific to a given molecule and thus can serve as good \textit{spectral fingerprints} \cite{haas2016advances}. 

However, \textit{near}-IR spectroscopy has many advantages as an experimental technique, advantages that often make it more desirable than mid-IR for detection applications \cite{huck2006near,bec2022silico}. In addition to being a low-cost and high-speed approach, weaker absorption means that it can be applied to much larger samples, as it is able to penetrate into the bulk; for the same reason, NIRS is applicable to a much wider variety of phases of matter \cite{siesler2008near,bik2020lipid}. At the same time, the ease of miniaturization of devices operating in the NIR part of the spectrum allows for truly portable detectors \cite{alcala2013qualitative,wiedemair2018application,adegoke2021near} -- most recently even reaching an individual consumer though cheap mobile phone attachments \cite{klakegg2016instrumenting,watanabe2016development,huang2021applications}. On the instrumentation side, the telecommunications industry has commercialized many of the technologies that enabled cheap and reliable NIRS -- chiefly among those the ability to use fiber optic cables for optic light transmission \cite{martin2002near}, an approach off-limits to mid-IR due to high signal attenuation in fiber optic in the associated spectral region. Finally, since very little or no sample preparation is required during spectrum acquisition \cite{ozaki2018near,sulub2007spectral}, the method is uniquely non-destructive and non-invasive. 

All these advantages make NIRS a desirable approach for chemical detection. Unfortunately, unlike their mid-IR counterparts, NIR spectra typically feature broad distributions of intensity spread across many wavelengths, rather than sharp isolated features that would make it easy to identify specific molecules by a handful of key wavelengths. While the spectra are still unique to a given molecule, a correct identification requires taking into account the global intensity pattern rather than a few isolated signature peaks \cite{sulub2007spectral,palafox2017computational,bec2019breakthrough,ozaki2021near}. 

What this means in practice is that identification via \textit{direct comparison} to stored spectra is no longer possible. Instead, it becomes necessary to \textit{train a statistical model} to detect a given molecule using its spectrum. A typical workflow for NIR-based chemical detection consists of the steps shown in \cref{fig:workflow} without the quantum simulation of the spectrum and associated dataset augmentation. Training statistical models requires large amounts of data to achieve high accuracy and transferability between measurement apparatuses and experimental contexts. Collecting a sufficient amount of data for model training is often expensive or otherwise challenging. This could be due to working with dangerous chemicals (e.g. toxic \cite{tan2019scheme} or explosive \cite{van2023rapid}), the conformers being hard to separate in the laboratory \cite{grabska2017temperature,sibert2023modeling}, or simply due to the large amount of data needed in the first place, for example when working with natural products \cite{nagy2022quality}.

Simulations of NIR spectra \cite{bec2021current,barone2021computational,ozaki2021advances}, a source of ``synthetic data'', may help address these limitations \cite{westad2008incorporating,bec2018nir,bec2019simulated,ozaki2021advances}. First, augmenting experimental datasets with simulated data could significantly alleviate the costs associated with data acquisition, and speed up the deployment process. Simulated data are also inherently safe to procure, removing any security concerns. Second, simulations provide additional information above and beyond the spectrum itself, for instance access to the underlying states and their symmetries, ability to vary geometry, environment effects, and so on. This information may be used to improve the sensitivity and specificity of statistical models, for example by selecting the key wavelengths most correlated with the property of interest on physical grounds, rather than purely statistically. And finally, this same information may also be used to aid in interpreting the results of a statistical model, building spectrum-structure relationships, thus enhancing reliability, trust, and transferability of the model \cite{westad2008incorporating, bec2018nir, bec2022silico}. 

\subsection{Classical simulation of NIRS}
    
Unlike in electronic structure, computational chemistry simulations methods for vibrational problems have received comparatively less attention. Most of the techniques from the former have found their counterparts in the latter: starting with the mean-field-like \gls{VSCF} \cite{vscf_1,vscf_2,vscf_3,vscf_4}, and continuing up the accuracy and complexity scale to standard vibrational perturbation theory (VPT) \cite{nielsen1951vibration,clabo1988systematic,lutz2014reproducible} and its many variants such as deperturbed and generalized flavours \cite{barone2005anharmonic,bloino2016aiming,krasnoshchekov2015ab,grabska2018nir}, vibrational coupled cluster (VCC) \cite{christiansen,christiansen2004vibrational,seidler2009automatic,seidler2011vibrational,faucheaux2015higher,hansen2020extended}, the already mentioned vibrational configuration interaction (VCI) \cite{fujisaki2007quantum,seidler2010vibrational,oschetzki2013azidoacetylene,schroder2022vibrational}, as well as the very recently introduced vibrational density-matrix renormalization group (vDMRG) \cite{baiardi2017vibrational,glaser2023flexible} and various selective or adaptive vibrational configuration interaction approaches \cite{scribano2008iterative,neff2009toward,fetherolf2021vibrational,schroder2021incremental}. 

In practice, the \gls{VSCF} approach is mostly helpful for molecules with weak anharmonicities in the PES, and is rarely used by itself in the strongly anharmonic NIR region \cite{roy2013vibrational,lutz2014reproducible,ozaki2021advances}. However, much like its Hartree-Fock counterpart, it serves as a starting point for all the higher-complexity methods, being a source of high-quality single-mode basis wavefunctions (called \textit{modals}) -- linear combinations of canonical harmonic oscillator modes that best capture anharmonicity effects in a mean-field way. 

The VCC and VCI approaches with a sufficiently high allowed excitation order (typically at least quadruple, e.g. VCI-SDTQ), can usually provide the accuracy needed in chemical detection applications, yielding spectra that match well to experiment \cite{oschetzki2013azidoacetylene,samsonyuk2013configuration,ozaki2021advances}. However, their prohibitive cost generally precludes their application to molecules with more than a dozen atoms, leading to them not being very widely used in practice. Vibrational perturbation theory is a middle ground, combining relative speed of execution with reasonable accuracy. But the technique is far from universal, requiring fine-tuning as it suffers from near-degeneracies (leading to the proliferation of approaches to tackling those \cite{barone2005anharmonic,bloino2016aiming,krasnoshchekov2015ab,grabska2018nir}), which become only more commonplace with increasing system size. While several methods have been developed to isolate and tackle such degeneracies, there are still numerous examples of where the VPT-produced spectrum differs significantly from the experimental one \cite{bec2018spectra,piccardo2015generalized,krasnoshchekov2015nonempirical,grabska2021theoretical}. 

Finally, for completeness we mention here the very recently introduced vDMRG \cite{baiardi2017vibrational,glaser2023flexible} and selective VCI \cite{scribano2008iterative,neff2009toward,fetherolf2021vibrational,schroder2021incremental} approaches. They have shown a lot of promise, especially when it comes to the mid-IR: however, these methods are yet to be tested more broadly, especially in the strongly-anharmonic NIR region. While the electronic counterparts of vDMRG and selective VCI are state-of-the-art for predicting electronic ground-state properties, they continue to face challenges when it comes to computing spectra: this manifests as limitations in terms of achievable system sizes. These tend to be especially severe for highly-excited states -- exactly the ones making up the NIR spectrum. 

Limitations of existing methods have motivated a few researchers to consider quantum approaches. This has led to intriguing arguments for why quantum advantage might manifest sooner in the vibrational structure problem rather than in electronic structure \cite{sawaya2021near}. There has also been prior work estimating the cost of computing vibrational spectra on fault-tolerant hardware, done on the example of a series of polymers of increasing size \cite{ibm_vibrational}. However, this resource estimate was very high, casting doubt on whether a quantum approach to vibrational problems would be worthwhile. We also note that an approach has been proposed for using the variational quantum eigensolver for the ground-state vibrational problem \cite{vqe_vib}. However, besides the convergence problems outlined in that work, it does not seem likely that the variational quantum eigensolver can be efficiently used for the spectroscopic problem where a large collection of excited states is required.

Motivated by this inability of existing classical to provide affordable high-quality training data for NIR-based chemical detection, in the rest of this paper we lay out a highly-optimized quantum algorithm for computing vibrational spectra of molecules. \cref{app:scaling} shows a deduction of the scaling of both our algorithm and VCI-SDTQ with respect to the number of modes $M$. We show how our algorithm scales as $\mathcal{O}(M^2)$, in contrast to the $\mathcal{O}(M^{12})$ scaling of VCI-SDQT. The latter comes from the $\mathcal{O}(M^4)$ states considered by VCI-SDTQ alongside the well-known $\mathcal{O}(N^3)$ scaling of matrix diagonalization for a matrix of size $N$, which here corresponds to $M^4$. This result, alongside the modest circuit sizes required for small molecules, present an area where early fault-tolerant quantum computers could unlock the full potential of NIRS-based characterization across many industries, which has been previously not possible due to the prohibitively high cost of performing these simulations on classical hardware. 
    
\section{Quantum algorithm for vibrational spectroscopy}
\label{sec:algorithm}

\begin{figure*}
    \centering
    \includegraphics[width=1.0\linewidth]{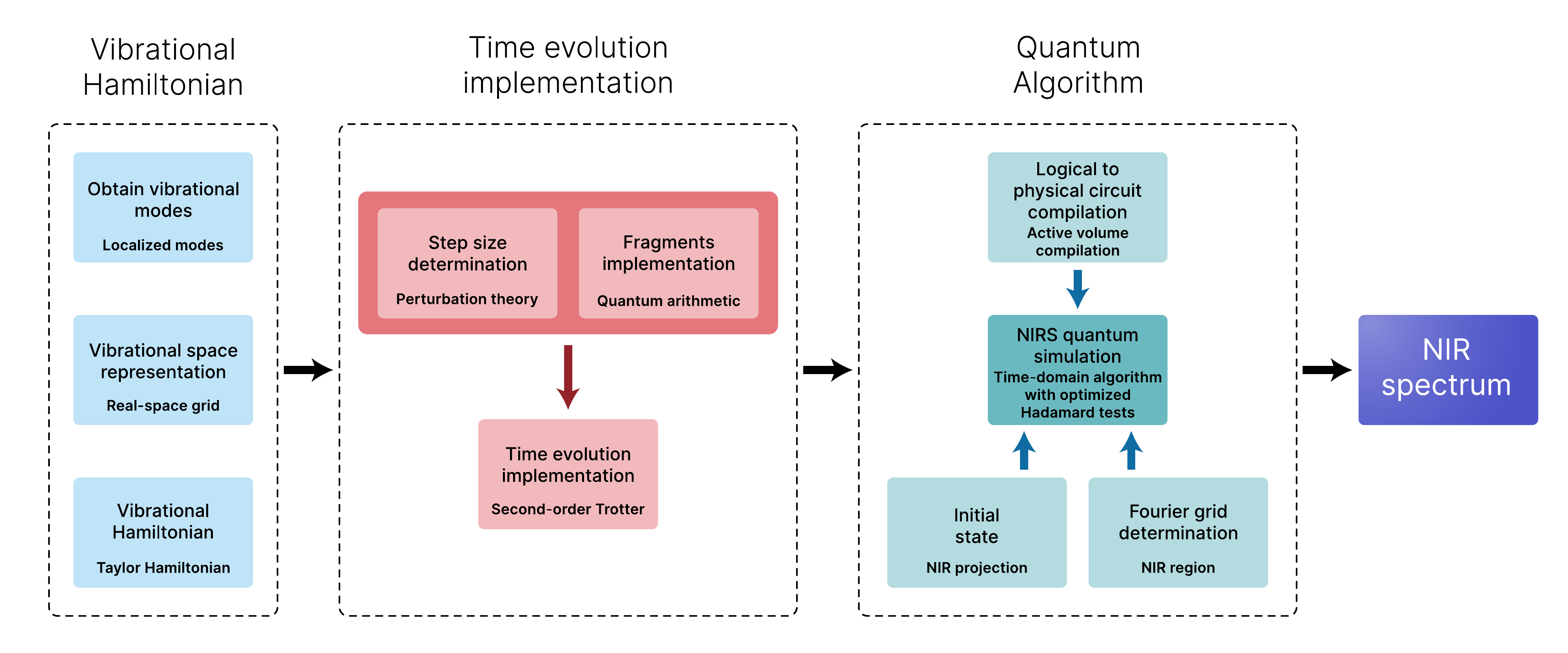}
    \caption{Workflow description for obtaining NIR spectrum on a quantum computer.  Titles in bold inside sub-blocks constitute the task, with underlying text describing the associated solution. Each of the three main blocks is described in a respective subsection of \cref{sec:algorithm}.}
    \label{fig:algo_hero}
\end{figure*}

We now provide a full description of the quantum algorithm we propose for obtaining the NIR absorption spectrum in \cref{eq:absorption}, with a summarized description presented in \cref{fig:algo_hero}. When performed on an appropriate initial state, the quantum algorithm recovers not only eigenvalues of a given Hamiltonian, but directly samples the absorption spectrum while also incorporating broadening effects \cite{qpe_xas,qpe_gqpe}. A deeper discussion of how our algorithm is connected to spectroscopy and \gls{QPE} is presented in Ref.~\cite{qpe_gqpe}, showing how to simulate linear and non-linear spectroscopic experiments on a quantum computer \cite{qpe_gqpe}. This algorithm can be performed directly in frequency space or with a time-domain approach that performs the Fourier transform on a classical computer \cite{qpe_lin_tong,qpe_cs,qpe_qmegs}. In this work we chose to use a time-domain implementation. An extensive optimization of all the elements that go into the algorithm was performed, allowing us to achieve a cost reduction by multiple orders of magnitude with respect to previous quantum approaches simulating vibrational systems \cite{ibm_vibrational}.  \\

This section is organized as follows. We first explain how to construct vibrational Hamiltonians in \cref{subsec:ham}. We then describe the quantum algorithm for obtaining NIR spectra in detail in \cref{subsec:qpe}. Finally, we explain all the details for implementing the time-evolution operator in \cref{subsec:time_evol}, noting that implementation of this operator accounts for most of the cost of the quantum algorithm.

\subsection{Vibrational Hamiltonian}
\label{subsec:ham}

In this work, we choose to employ the vibrational Hamiltonian representation based on the Taylor expansion of the PES; we will refer to this as the \textit{Taylor form} for brevity. The full details of how this Hamiltonian is derived from first-principles are presented in \cref{app:taylor}. This choice allows us to work with a real-space representation, which is one of the reasons for improved cost estimates for our algorithm. At the end of this section, we mention the alternative representations and briefly comment on our choice.  

The starting point is the set of normal coordinates $\{ q_1,\cdots,q_M\}$, expressed in natural units, where each $q_j$ is a scalar associated with a collective displacement of the atoms. Such coordinates are normally obtained by diagonalizing the Hessian of the PES near the equilibrium geometry. Then the harmonic component of the Hamiltonian is written as
\begin{equation}
    \hat H_{\rm harm} = \frac{1}{2}\sum_{j=1}^M \hbar\omega_j \left(\hat p_j^2 + \hat q_j^2\right),
\end{equation}
where $\hat p_j$ is the momentum operator associated with $\hat q_j$, $\omega_j$ is the harmonic frequency of mode $j$, and $M=3N_{\rm atoms} - 6$ is the number of vibrational modes, with $M=3N_{\rm atoms} - 5$ for linear molecules. The position and momentum operators follow the canonical commutation relations $[\hat q_j,\hat q_k] = [\hat p_j,\hat p_k] = 0$ and $[\hat q_j,\hat p_k] = i\hbar\delta_{jk}$. The full vibrational Hamiltonian $\hat H$ is then written as
\begin{equation}
    \hat H = \hat H_{\rm harm} + \hat H_{\rm anh},
\end{equation}
where the anharmonic component can be expressed by a Taylor expansion of the PES with respect to the normal coordinates
\begin{equation}
    \hat H_{\rm anh} = \sum_{i\geq j\geq k}
    \Phi_{ijk}^{(3)} \hat q_i\hat q_j\hat q_k + \sum_{i\geq j\geq k\geq l} \Phi_{ijkl}^{(4)} \hat q_i\hat q_j\hat q_k \hat q_l + \cdots. \label{eq:taylor_ham} 
\end{equation} 
The Taylor coefficients can be obtained from doing a fitting of the PES; see \cref{app:taylor} for details. In the rest of this manuscript, we will refer to two attributes of a particular vibrational Hamiltonian: its maximum Taylor expansion order, which corresponds to the largest sum-total monomial power appearing in the expansion; and its $n$-mode character, where $n$ is the maximum number of distinct modes appearing together in a monomial. As an example, a Hamiltonian $\hat H = \hat q_1\hat q_2\hat q_3\hat q_4$ would have Taylor order four and be $4$-mode, while $\hat H = \hat q_1^2\hat q_2\hat q_3$ would also have Taylor order four while only being $3$-mode. In practice one often restricts the allowed inter-mode interactions to $3$- or even $2$-mode, while maintaining Taylor order of four: this corresponds to setting coefficients such as $\Phi_{ijkl}$ or $\Phi_{ijkk}$ to zero.  \\

The most expensive subroutine in our algorithm is the time-evolution operator $e^{-i\hat H t}$. The cost of this operation is directly related to the number of terms in the Taylor expansion in \cref{eq:taylor_ham}. Any techniques that allows us to reduce this number of terms while not hindering accuracy are particularly useful for reducing the overall cost of the quantum algorithm. In this work, we use the mode localization approach \cite{mode_loc_1,mode_loc_2,mode_loc_3,mode_loc_4}, which effectively makes the Hamiltonian more sparse and lowers the overall implementation cost. As discussed in \cref{app:mode_loc}, the usage of mode localization techniques has been shown to greatly reduce the required number of inter-mode interactions \cite{mode_loc_1}, with the example of the ethane C$_2$H$_4$ Hamiltonian yielding the same accuracy using a $2$-mode representation with localized modes when compared to using a $4$-mode expansion with (canonical) normal modes. By analogy to the use of localized orbitals in electronic structure calculations, mode localization redefines the vibrational modes $\hat q_i$'s so as to make the vibrations assume a more local character in the molecule. This is in contrast to the normal modes of vibration that are typically described by collective displacements of all atoms in a molecule. Mode localization yields a Hamiltonian of the form
\begin{align} 
    \hat H &= \sum_{i\geq j} \left( K_{ij} \hat p_i \hat p_j + \Phi_{ij}^{(2)} \hat q_i \hat q_j \right) + \sum_{i\geq j\geq k} \Phi_{ijk}^{(3)} \hat q_i \hat q_j \hat q_k \nonumber \\
    &\ \ + \sum_{i\geq j\geq k\geq l} \Phi_{ijkl}^{(4)} \hat q_i \hat q_j \hat q_k \hat q_l  + \cdots,
    \label{eq:loc_ham}
\end{align}
the main difference being the introduction of non-diagonal momentum terms. The coordinates $\hat q_i$ and coefficients $\Phi$ have been transformed, but for brevity we will use the same symbols and from now on refer exclusively to localized modes, unless explicitly stated otherwise. A more thorough discussion of how the modes are localized and the benefits of using this technique can be seen in \cref{app:mode_loc}. \\

We now comment on our choice of the Taylor form. Two alternative representations of the Hamiltonian are known.
The most common representation is to pass from position and momentum operators $\hat q, \hat p$ to the bosonic ladder operators $b, b^\dagger$, re-expressing the higher-order potential energy surface terms like $\hat q_i \hat q_j \hat q_k$ as powers of these creation and annihilation operators \cite{baiardi2017vibrational}. We refer to this as the bosonic representation. A different, relatively recent approach is to use the so-called Christiansen representation (sometimes also somewhat vaguely called the occupation-number representation) first introduced in Ref.~\cite{christiansen}, where the bosonic ladder of excitations for each mode is truncated and a new register is associated to each modal. 

However, using the real-space representation has two concrete benefits over these second-quantized approaches. First, the real-space form discretizes each vibrational mode with exponential accuracy with respect to the used number of qubits, which alongside the compact Taylor representation of the Hamiltonian yields significantly fewer terms when compared to using a second-quantized approach. More importantly, the real-space form allows the use of highly-optimized quantum arithmetic primitives (addition and multiplication) together with the phase gradient trick to implement time-evolution \cite{vibronic,gidney2018halving,su2021fault,quantum_arithmetic}, resulting in a cheaper quantum algorithm overall. More details on the comparison between the three representations are in \cref{app:cf_reps}, and we describe how to use quantum arithmetic to implement time-evolution by the Hamiltonian in the Taylor form in \cref{subsec:time_evol}. \\

\begin{figure}
    \centering
    \includegraphics[width=1\linewidth]{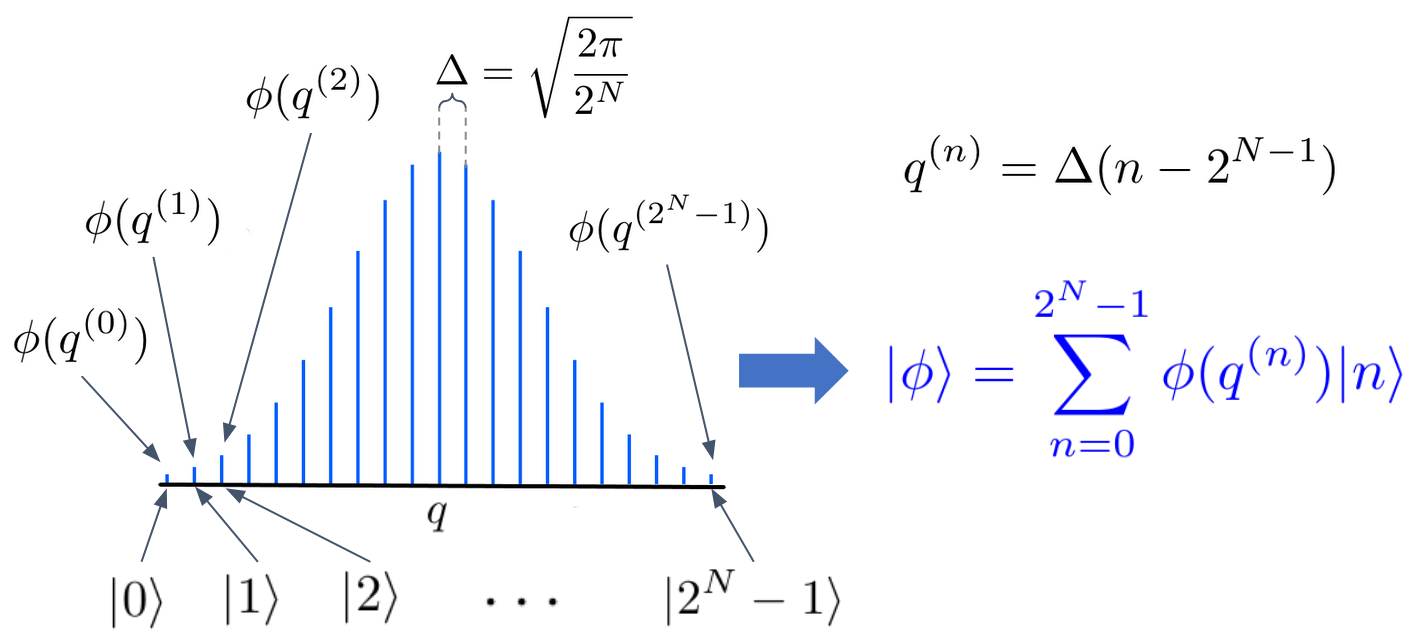}
    \caption{Real-space representation \cite{macridin_1,macridin_2} of single vibrational mode $q$. Each of the $2^N$ integers associated with the $N$-qubit computational basis states are associated with a different point in the grid discretizing $q$. The wavefunction $\ket{\phi}$ will then be represented by a coherent superposition over computational basis states with associated coefficients $\phi(q^{(n)})$, namely $\ket{\phi} = \sum_n \phi(q^{(n)}) \ket{q^{(n)}}$.}
    \label{fig:grid}
\end{figure}

Once we have obtained the Taylor expansion constants in \cref{eq:loc_ham}, we need to define a qubit-based representation of the vibrational space. In this work we adopt the real-space representation as introduced in Refs.~\cite{macridin_1,macridin_2} and also used for vibrational degrees of freedom in Ref.~\cite{vibronic}. The basic idea is to have a qubit register consisting of $N$ qubits for each vibrational mode. For a given mode $q$, each of the computational basis states then encodes a point in a uniform grid that is associated to the discretization of $q$ using $2^N$ grid points as shown in \cref{fig:grid}. From this we have $2^N$ different computational basis states for each vibrational mode: the $n$-th point in the grid for a vibrational mode $q$ has an associated computational basis state such that 
\begin{equation} \label{eq:qn}
    \hat q \ket{n} = \Delta(n - 2^{N-1}) \ket{n},
\end{equation}
where we have defined the constant $\Delta\equiv \sqrt{2\pi / 2^N}$ with an associated total grid width of $\sqrt{\pi 2^{N+1}}$. As thoroughly discussed in Refs.~\cite{macridin_1,macridin_2}, the number of qubits $N$ required for achieving a faithful representation of vibrational excitations consisting of up to $N$ quanta in this grid scales logarithmically with respect to $N$. This exponential accuracy with respect to $N$ can be understood via the Nyquist-Shannon sampling theorem, and effectively allows for accurate encoding of the vibrational space with a small number of qubits. Applying the \gls{QFT} on a given mode's qubit register is then simply a change of basis between the position and momentum representations: the transformation has the form
\begin{equation}
    \hat p = \mathtt{QFT}^\dagger \cdot \hat X_{s} \cdot \hat q \cdot \hat X_{s} \cdot \mathtt{QFT},
\end{equation}
where $\hat X_{s}$ corresponds to an $X$ gate on the most significant qubit of the register encoding the vibrational mode $q$. This $X$ gate effectively performs a ``shift'' in the same way as usually done for the discrete Fourier transform in classical computing, which effectively centers the Fourier transform in some range $[-a,a)$ instead of $[0,2a)$. As such we will refer to the combination $\hat X_{s} \cdot \mathtt{QFT}$ simply as the shifted \gls{QFT}.

\subsection{Fourier-based algorithm for NIRS} \label{subsec:qpe}
In this section we discuss the quantum algorithm for obtaining the NIR spectrum, alongside its relationship with Fourier series, and the associated optimizations that were done in this work.

\subsubsection{Fourier representation and initial state filtering} \label{subsubsec:fourier}

Having obtained the vibrational Hamiltonian, we can use it to obtain its absorption spectrum as shown in \cref{eq:absorption}, namely
\begin{equation*} 
    \sigma_A(\omega) \propto \omega \sum_{\rho=x,y,z}\sum_{f} |\bra{E_f} \hat\mu_\rho \ket{E_0}|^2 \mathcal{L}_\eta(\omega-E_f+E_0).
\end{equation*}

As discussed in Refs.~\cite{qpe_gqpe,qpe_xas}, when using an appropriate input state \gls{QPE} directly samples frequencies from the associated absorption spectrum. When applied to the normalized initial state
\begin{equation} \label{eq:init_state}
    \ket{\psi_\rho} = \frac{\hat\mu_\rho \ket{E_0}}{|\mu_\rho|},
\end{equation}
where we use the short-hand $|\mu_\rho|=||\hat\mu_\rho\ket{E_0}||$,
the canonical \gls{QPE} procedure returns the frequency-dependent distribution
\begin{equation} \label{eq:qpe_out}
    P_\rho(\omega) = \frac{1}{|\mu_\rho|^2}\sum_{f} |\bra{E_f} \hat\mu_\rho \ket{E_0}|^2 \mathcal{L}(\omega - E_f).
\end{equation}
The line shape function $\mathcal{L}(\omega)$ is linked to the initial preparation of the time/frequency register, as explained in Ref.~\cite{qpe_gqpe}.  \\

We will now show how this quantity can be expressed as a Fourier series. Before the addition of the line shape, we can write the similar Fourier integral
\begin{align} \label{eq:fourier_integral}
    Z_\rho(\omega) &\equiv \int_{-\infty}^\infty dt e^{i\omega t} \bra{\psi_\rho}e^{-i\hat H t} \ket{\psi_\rho} \\
    &= 2\pi \sum_f |\bra{E_f}\hat\mu_\rho\ket{E_0}|^2 \delta(\omega - E_f).
\end{align}
From this we can obtain the target distribution as
\begin{equation}
    P_{\rho}(\omega) = \frac{1}{2\pi |\mu_\rho|^2} Z_\rho(\omega) *\mathcal{L}(\omega),
\end{equation}
where $*$ defines the convolution operation. Using the convolution theorem, and considering that we are targeting a Lorentzian line shape such that $\mathcal{L}(\omega) = \eta / (\omega^2 + \eta^2)$ with associated Fourier conjugate $e^{-\eta|t|}/2$, we arrive to
\begin{equation}
    P_{\rho}(\omega) = \frac{1}{2\pi |\mu_\rho|^2} \int_{-\infty}^\infty dt e^{i\omega t} \frac{e^{-\eta |t|} }{2}\bra{\psi_\rho}e^{-i\hat H t}\ket{\psi_\rho}.
\end{equation}
The next step is to approximate the integral on the right-hand side to some accuracy $\epsilon$ by truncating the infinite time as $\int_{-\infty}^\infty\rightarrow\int_{-T_{\rm max}}^{T_{\rm max}}$, which as shown in \cref{app:max_time} can be done by choosing
\begin{equation} \label{eq:max_time}
    T_{\rm max} = \frac{1}{\eta} \log \frac{1}{\epsilon}.
\end{equation}
This finite integral can now be expressed using a Fourier series. Noting that $P_\rho(\omega)$ is localized inside of some spectral range $[\omega_{\rm min}, \omega_{\rm max}]$, we can use a finite Fourier series to expand $P_\rho(\omega)$, effectively making it periodic outside of this range. Defining the spectral width as $\Omega = \omega_{\rm max} - \omega_{\rm min}$, this effectively defines a Fourier grid for a discrete Fourier transform. We thus obtain the Fourier series approximation of the target distribution
\begin{equation} \label{eq:spectrum_from_ks}
    P_\rho(\omega) \approx \frac{1}{\Omega|\mu_\rho|^2}\left[ \frac{1}{2}+\sum_{k=1}^{k_{\rm max}} \operatorname{Re}\left( p_\rho[k] e^{2\pi ik\omega/\Omega}  \right)\right]
\end{equation}
where the number of Fourier coefficients is 
\begin{equation} \label{eq:kmax}
k_{\rm max}=\lceil \Omega \cdot T_{\rm max}\rceil    
\end{equation}
and we defined the Fourier components
\begin{equation} \label{eq:fourier_component}
    p_\rho[k] = e^{-2\pi\eta k/\Omega} \bra{\psi_\rho} e^{-2\pi i\hat H k/\Omega} \ket{\psi_\rho}.
\end{equation}
The unitary that is entering the expectation value is $e^{-2\pi i\hat H k/\Omega} $, which can also be thought as rescaling the Hamiltonian $\hat H\rightarrow \hat H /\Omega$ to effectively estimate the phase of $\hat H/\Omega$. Typically a rescaling factor $\Omega\sim\mathcal{O}(||\hat H||^{-1})$ is used in order to keep the spectrum in the $[-1,1]$ range and avoid aliasing in the Fourier transform that comes from making the $P_\rho(\omega)$ function periodic when approximating it as a Fourier series. 

\begin{figure}
    \centering
    \includegraphics[width=1\linewidth]{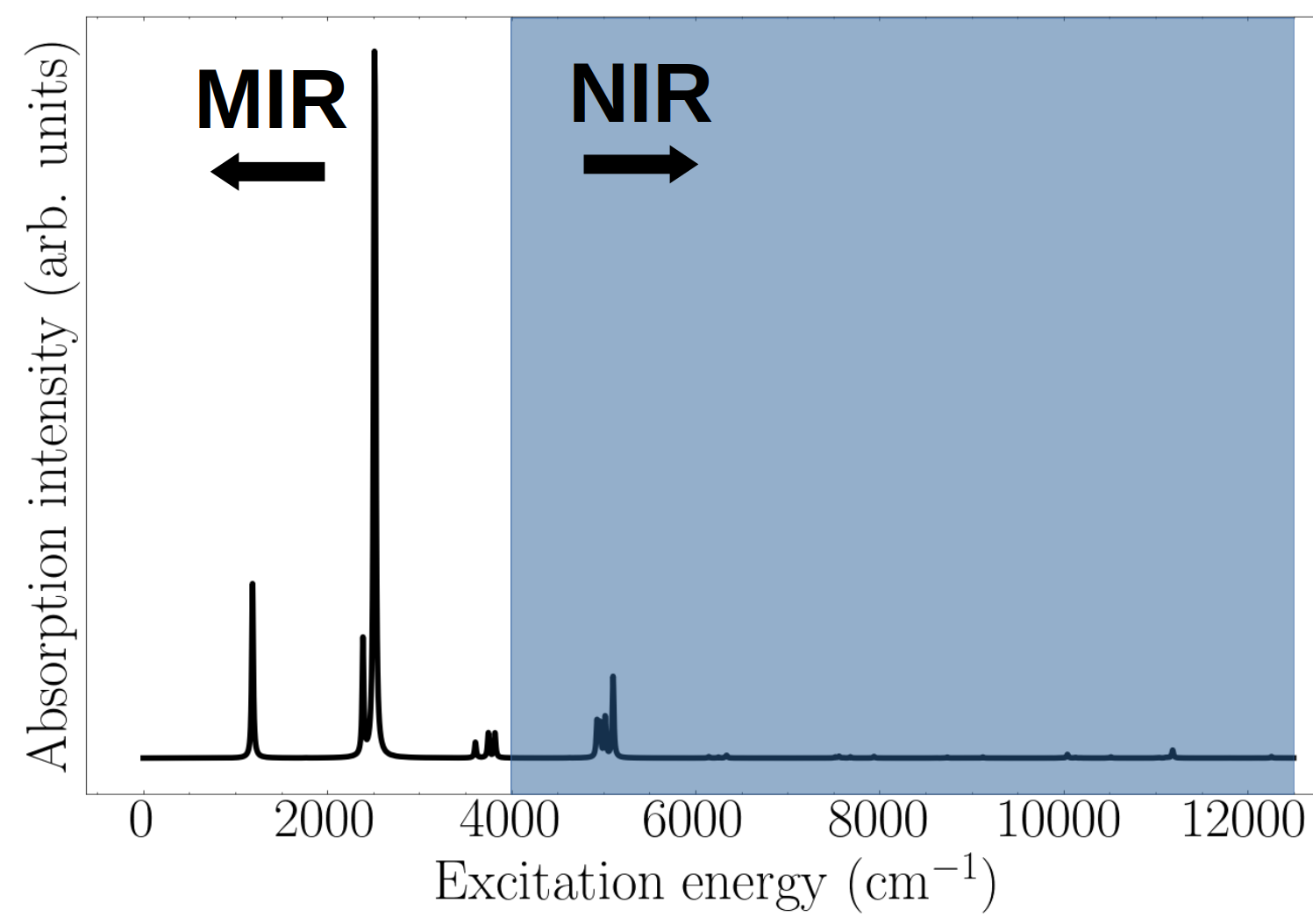}
    \caption{Simulated spectrum of hydrogen sulfide H$_2$S molecule. Note how the amplitudes of the \textit{near}-IR spectrum are significantly smaller than those of the \textit{mid}-IR: the former come from overtones and combination bands, having a strong anharmonic character that is classically challenging to simulate.}
    \label{fig:h2s_full}
\end{figure}

However, in this work we are focusing in the NIR spectrum, which consists of the range of frequencies between $4,000$ and $12,500$ cm$^{-1}$. As seen in \cref{fig:h2s_full}, the mid-IR region compromises the largest absorption peaks, while the NIR region has peaks coming from overtones and combination bands that quickly decay as the excitation energy increases, while no significant vibrational excitations happen beyond this region. Since our interest is in the NIR region, it becomes unnecessary to simulate the mid-IR portion. Still, if we choose the Fourier frequency window to be only in the NIR range, the non-zero spectrum coming from mid-IR would appear as an aliasing effect due to the periodic nature of the Fourier series. In order to avoid simulating any peaks from this region, we propose a filtering procedure over the initial state $\ket{\psi_\rho}$, which can be written in the eigenstate basis as
\begin{equation} \label{eq:init_eigen}
    \ket{\psi_\rho} = \sum_k a_k^{(\rho)} \ket{E_k}.
\end{equation}
The idea is to set to zero all $a_k^{(\rho)}$ with associated energies $E_k$'s outside of the NIR range. This can be written as the condition $\omega_{k0}<4,000\textrm{ cm}^{-1}$, where we have defined the excitation energy $\omega_{k0} \equiv E_k - E_0$. However, in practice we obviously do not have access to the exact eigenstates and energies. To circumvent this we start by considering some approximate eigenstates $\ket{\tilde E_k}$, which in this work we obtain from the mean-field method VSCF \cite{vscf_1,vscf_2,vscf_3,vscf_4}. We then expand the wavefunction in this basis as $\ket{\psi_\rho} = \sum_k \tilde a_{k}^{(\rho)} \ket{\tilde E_k}$. Due to the approximate nature of these eigenstates and in order to avoid removing components that could appear in the NIR spectrum, we used a padding of $250{\rm \ cm^{-1}}$, noting that VSCF energies typically approximate the exact energies in the mid-IR region within $\sim 100$ cm$^{-1}$ \cite{beyond_vscf}. We thus chose an energy cutoff value of $W=3,750$ cm$^{-1}$. By considering the projector
\begin{equation}
    \hat\Pi_W = \sum_{k | \tilde\omega_{k0} \geq W} \ket{\tilde E_k}\bra{\tilde E_k},
\end{equation}
we can pass the (renormalized) initial state $\hat\Pi_W \ket{\psi_\rho}$ to our algorithm to focus on the NIR region.  From this consideration, we chose to simulate the window in the spectral range between $\omega_{\rm min} = 3,500\textrm{ cm}^{-1}$ and $\omega_{\rm max} = 12,500$ cm$^{-1}$ as to guarantee that no absorption line coming from the mid-IR region appears as an aliasing component in our recovered spectrum. This yields the spectral width
\begin{equation}
    \Omega = 9,000\textrm{ cm}^{-1}.
\end{equation}
Besides allowing us to simulate only the NIR region with the smaller $\Omega$, which translates to smaller $k_{\rm max}$ and maximum evolution times, the removal of these states in the mid-IR entails a renormalization of the initial wavefunction by a factor of $|\hat\Pi_W\ket{\psi_\rho}|$. The spectrum obtained with the initial wavefunction $\hat\Pi_W\ket{\psi_\rho}/|\hat\Pi_W\ket{\psi_\rho}|$ carries a division by $||\hat\Pi_W\ket{\psi_\rho}||^2$ when compared to the one obtained with $\ket{\psi_\rho}$ in \cref{eq:spectrum_from_ks}. The overall accuracy in the NIR region is thus boosted by a factor of $||\hat\Pi_W\ket{\psi_\rho}||^{-2}$. This manifests as a large improvement in the accuracy requirements by around two to six orders of magnitude: the $\hat\mu_\rho\ket{E_0}$ optical excitation is typically constituted by $90$ to $99.9\%$ of states coming from the mid-IR region. \\

Before continuing the discussion of the algorithm, we now explain why using a product formula implementation of the time-evolution operator is more attractive than using a qubitization-based framework \cite{qubitization,qrom,mtd,walk_based_qpe}. Of particular interest for this comparison is the ability to focus on the NIR spectral region in our product-formula-based implementation. Being able to focus in some particular spectral region cannot be easily done when performing the so-called ``walk-based \gls{QPE}'' approach \cite{qrom}, noting that walk-based \gls{QPE} constitutes a more efficient approach for qubitization-based \gls{QPE} when compared to direct implementation of the time-evolution operator through the Jacobi-Anger expansion \cite{qubitization}. Qubitization necessitates the rescaling of the entire spectrum of $\hat H$ to the range $[-1,1]$ through the appearance of the 1-norm $\lambda$, where the operator $\hat H/\lambda$ is effectively implemented as to keep the block-encoding unitary \cite{LCU4}. However, in practice most of the absorption in this rescaled spectrum will be trivially zero. This happens since the representation used for the Hamiltonian will have many states that are never excited but with extremely high energies. An example of such a state would correspond to having all the amplitude of a given state in the real-space representation appearing only in the borders of the grid. From this we can assert that as the system size grows and more of these states become available, $\lambda$ will quickly increase, whereas our chosen $\Omega$ has a constant value of $9,000\textrm{ cm}^{-1}$ associated with the NIR region, which is the only region where there will be some absorption after the state projection procedure. Rescaling by $\Omega$ instead of $\lambda$ then enables a vastly increased resolution in the energy x-axis. This consideration makes product-formula-based implementations of \gls{QPE}-based algorithms for spectroscopy particularly attractive.\\ 

\subsubsection{Time-domain quantum algorithm} \label{subsubsec:time_domain}

Having shown how to express the target distribution as a Fourier series with associated coefficients $p_\rho[k]$ in \cref{eq:fourier_component}, these now need to be determined for each of $k=1,\cdots,k_{\max}$. The $e^{-2\pi \eta k/\Omega}$ factor makes contributions from large $k$'s vanishingly small, which in turn means that there is no need to determine the associated $p_\rho[k]$'s to the same high accuracy as those coming from smaller $k$'s. Overall, we will allocate a number of shots proportional to $e^{-2\pi\eta k/\Omega}$ to determine the associated $p_\rho[k]$, which as seen below allows us to seamlessly incorporate the Lorentzian line shape while also having a fixed accuracy for each $p_\rho[k]$ component.

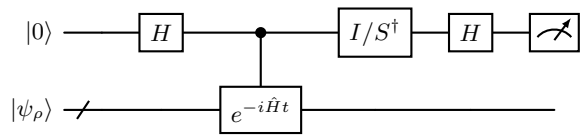
\begin{figure}
    \centering
    \begin{quantikz}[row sep=0.4cm]
		\lstick{$\ket{0}$} && \gate{H} & \ctrl{1} & \gate{I/S^\dagger} & \gate{H} & \meter{} \\ 
		\lstick{$\ket{\psi_\rho}$} & \qwbundle{} && \gate{e^{-i\hat H t}} &&&
	\end{quantikz}
    \caption{Hadamard test circuit for a given evolution time $t$. By obtaining expectation values over a range of times and performing the Fourier transform on classical hardware \cref{eq:qpe_out} is recovered.}
    \label{fig:hadamard_qpe}
\end{figure}

The key component of the time-domain formulation of the algorithm is the expectation value of the self-correlation function$\bra{\psi_\rho}e^{-i\hat H t} \ket{\psi_\rho}$. Shown in \cref{fig:hadamard_qpe} is the Hadamard test quantum circuit for recovering this quantity through the implementation of the time-evolution for a given time $t$: the ancilla qubit encodes the expectation value of interest as
\begin{equation} \label{eq:hadamard_expectation}
    p_0(t)-p_1(t) = \operatorname{Re}\left[{\bra{\psi_\rho} e^{-i\hat H t} \ket{\psi_\rho}}\right].
\end{equation}
The imaginary component can be obtained instead if the gate in the Hadamard test after the controlled time-evolution is $S^\dagger$ instead of $I$. Here $p_0(t)$ and $p_1(t)$ are the probabilities of observing the ancilla in state $0$ and $1$ respectively for the controlled time-evolution $e^{-i\hat H t}$. 

As shown in Refs.~\cite{qpe_qmegs,qpe_gqpe,qpe_lin_tong}, if we consider a total number of shots $\mathcal{S}$, we can then run the Hadamard test circuit in \cref{fig:hadamard_qpe} for a time $t=2\pi k/\Omega$ a total of 
\begin{equation} \label{eq:shot_normalization}
    \mathcal{N}_k = \mathcal{S}\mathcal{N} e^{-2\pi \eta k/\Omega}
\end{equation}
times for both the real and imaginary components, where $\mathcal{N} = \left[2\sum_{k=1}^{k_{\max} }e^{-2\pi \eta k/\Omega}\right]^{-1}$ is a normalization constant. The $m$-th pair of real/imaginary measurements of the ancilla qubit with associated $k=k_m$ then yields the numbers $x_\rho^{(m)}$ and $y_\rho^{(m)}$. Here we have assigned a value of $1(-1)$ to the random variables $x_\rho^{(m)}$ and $y_\rho^{(m)}$ when the associated measurement of the Hadamard test ancilla qubit corresponds to $0(1)$, which implies that $\bra{\psi_\rho}e^{-2\pi i\hat H k/\Omega}\ket{\psi_\rho} = \mathbb{E}\left[{x_\rho^{(m)} + iy_\rho^{(m)}}\right]$ . The non-uniform sampling over $k$'s then incorporates the line shape contribution to the spectrum, recovering the target spectrum as
\begin{align} \label{eq:shot_spectrum}
    P_\rho(\omega) &\approx \frac{1}{\Omega|\mu_\rho|^2} \left[\frac{1}{2} + \frac{1}{\mathcal{N}_0}\sum_{m=1}^{\mathcal{S}/2} \operatorname{Re}\left[\left(x_\rho^{(m)}+iy_\rho^{(m)}\right) e^{2\pi ik\omega/\Omega} \right]  \right] ,
\end{align}
where $\mathcal{N}_0$ corresponds to considering \cref{eq:shot_normalization} for $k=0$. To better understand the effect of the variable number of shots in the determination of the spectrum, we now provide a bound of the variance associated with the Fourier components of the spectrum [\cref{eq:fourier_component}], namely $p_\rho[k] = e^{-2\pi\eta k/\Omega} \cdot \mathbb{E}\left[x_\rho^{(m)} + iy_\rho^{(m)}\right]$. The variance after considering $\mathcal{N}_k$ samples of these random variables is
\begin{align}
    {\rm Var}(p_\rho[k]) &= \frac{1}{\mathcal{N}_k^2} [e^{-2\pi\eta k/\Omega}]^2 \cdot {\rm Var}(x_\rho^{(m)}+iy_\rho^{(m)}) \\
    &\leq \frac{1}{(\mathcal{SN})^2} \cdot \mathbb{E}\left[|x_\rho^{(m)}|^2 + |y_\rho^{(m)}|^2\right] \\
    &= \frac{2}{(\mathcal{SN})^2}.
\end{align}
From this analysis we can see how the $k$-dependent shot allocation in \cref{eq:shot_normalization} entails a fixed maximum variance over all the different Fourier components of the spectrum. \\

We note that some approaches have been proposed for post-processing the obtained spectrum in \cref{eq:shot_spectrum} by using, e.g., a matching pursuit algorithm \cite{qpe_qmegs}, which can be effectively used to remove the noise coming from finite sampling. Determination of $\mathcal{S}$ is purely a signal processing problem, corresponding to the accuracy that is required for the coefficients of a Fourier series to recover the associated signal to some accuracy, with the main quantities that determine the signal being the line shape width $\eta$ and the spectral range $\Omega$. In practice we use the quantity $\mathcal{S}_{\rm tot}=3\mathcal{S}$ to account for all three Cartesian components of the dipole appearing in the absorption spectrum. Practical choices of $\mathcal{S}_{\rm tot}$ are obtained heuristically in \cref{sec:benchmark}.

\subsubsection{Hadamard test optimizations} \label{subsubsec:hadamard}
Having shown how a measurement of the expectation values in \cref{eq:hadamard_expectation} through the Hadamard test can be used to recover the NIR spectrum, we now discuss two additional optimizations that effectively diminish the required evolution times when implementing $e^{-i\hat H t}$ and the required number of circuit repetitions \cite{new_xas}.  \\

\begin{figure*}
    \centering
       \begin{quantikz}[row sep=0.4cm]
       \lstick{$\ket{0}$} && \gate{H} & \ctrl{1} & \gate{I/S^\dagger} & \gate{H} &  \meter{} &  \setwiretype{n} &&& \\
		\lstick{$\ket{\vec 0}$} & \qwbundle{} & \gate{U_\rho} & \gate{U} & \gate{U_\rho^\dagger} && \meter{} 
	\end{quantikz}
    \caption{Double measurement Hadamard circuit which un-prepares the initial state and then measures the output of the Hadamard circuit, returning additional information about the expectation value $\bra{\psi_\rho}\hat U \ket{\psi_\rho}$.}
    \label{fig:double_measurement}
\end{figure*}
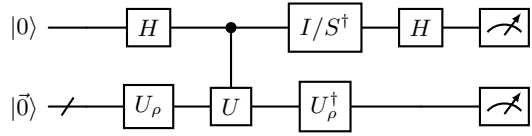

\paragraph{Double measurement method.}
The basic idea is to re-use the output state after applying the Hadamard test to gain additional information of the associated expectation value. After performing the Hadamard circuit in \cref{fig:hadamard_qpe}, by undoing the initial state preparation in the system register and measuring the system qubits we can increase the amount of information that is recovered from each Hadamard test. The circuit for performing this can be seen in \cref{fig:double_measurement}. For simplicity in this discussion we assume that the Hadamard test was ran obtaining the real part, noting that an analogue analysis can be done for the imaginary component. The full details of this procedure are shown in Ref.~\cite{new_xas}. \\

We start by noting that the output state of the Hadamard test corresponds to
\begin{equation}
    \ket{\Psi_\rho^\pm} =\frac{(\hat I \pm \hat U) \ket{\psi_\rho}}{\sqrt{\bra{\psi_\rho}(1\pm\hat U)^\dagger(1\pm\hat U)\ket{\psi_\rho}}},
\end{equation}
where the $\pm$ sign is determined by the outcome from measuring the Hadamard test ancilla qubit. After applying the hermitian conjugate of the initial state preparation $\hat U_\rho$ on the system, which acts as $\hat U_\rho\ket{\vec 0} = \ket{\psi_\rho}$, the probability of measuring all $0$'s in the system register then corresponds to
\begin{equation}
    p^{(\pm)}_{\vec 0} = \frac{|1\pm \bra{\psi_\rho}\hat U\ket{\psi_\rho}|^2}{2(1\pm {\rm Re}\bra{\psi_\rho}\hat U\ket{\psi_\rho})}.
\end{equation}
The full analysis of how this technique lowers the overall cost is presented in Ref.~\cite{new_xas}: the result is that the overall number of shots is reduced by a factor $\zeta \in [1,3.00]$, the factor being determined by the expectation value being measured -- with an average reduction of $\zeta=1.49$. \\

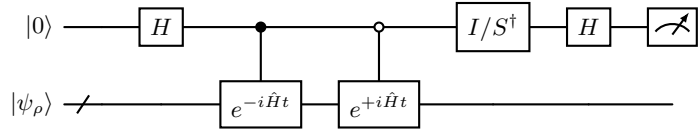
\begin{figure*}
    \centering
    \begin{quantikz}[row sep=0.4cm]
		\lstick{$\ket{0}$} && \gate{H} & \ctrl{1} & \octrl{1} & \gate{I/S^\dagger} & \gate{H} & \meter{} \\
		\lstick{$\ket{\psi_\rho}$} & \qwbundle{} && \gate{e^{-i\hat H t}} &  \gate{e^{+i\hat H t}}&&&
	\end{quantikz}
    \caption{Double phase circuit which effectively doubles the time-evolution length on the Hadamard test in \cref{fig:hadamard_qpe}. }
    \label{fig:double_phase}
\end{figure*}

\paragraph{Double phase method.}
The main idea of this method~\cite{new_xas,qrom,double_phase} is to change the controlled unitary that the Hadamard test implements as
\begin{equation} \label{eq:double_hadamard}
    \begin{bmatrix}
        \hat 1 & 0\\
        0 & e^{-i\hat H t}
    \end{bmatrix} \rightarrow
    \begin{bmatrix}
        e^{i\hat H t} & 0\\
        0 & e^{-i\hat H t}
    \end{bmatrix},
\end{equation}
as shown in \cref{fig:double_phase}. The benefit of doing this is that the ancilla qubit now encodes the expectation value of $e^{-2i\hat H t}$ instead of the $e^{-i\hat H t}$ term seen in \cref{eq:hadamard_expectation}, which implies recovering the expectation value for an evolution time that is twice as long. Moreover, implementing the unitary corresponding to \cref{eq:double_hadamard} can be done for an even lower cost than the controlled time-evolution in the original Hadamard test. To understand this improvement, we start by considering the original Hadamard test, for which controlling the time-evolution simply necessitates controlling the addition operation to the circuit shown in \cref{fig:poly_evolution}. As explained in \cref{subsec:time_evol}, the coefficients coming from the Taylor expansion, e.g. $\Phi_{ij}^{(2)}$, are encoded in a quantum register, with the associated $e^{i\Phi_{ij}^{(2)}\hat q_i\hat q_j}$ operator then implemented using the phase kickback technique \cite{phase_gradient}. Thus, instead of having to control the addition operation, the unitary in \cref{eq:double_hadamard} for the modified Hadamard test can be implemented by simply flipping the qubit that encodes the sign of the $\Phi_{ij}^{(2)}$ coefficient, which is done using a $\mathtt{CNOT}$ gate. When we also consider the uncomputation cost (see \cref{fig:poly_evolution}) the modified unitary in \cref{eq:double_hadamard} can be implemented with the same cost of the uncontrolled application of $e^{-i\hat H t}$, plus $2N_{\rm Taylor}$ $\mathtt{CNOT}$ gates, where $N_{\rm Taylor}$ is the total number of non-zero coefficients appearing in the Hamiltonian in \cref{eq:taylor_ham}. This basically allows to implement the required time-evolutions for half the cost. 

\subsubsection{Active volume compilation} \label{subsubsec:act_vol}

The active volume compilation technique was introduced in Ref.~\cite{active_volume}.  At an abstract algorithmic level, it can be thought of as reducing the time and resources needed to run quantum algorithms on a specific error-corrected hardware architecture. The basic idea is based on the observation that during typical computations, most of the qubits at any given time are idle. However, the computation cost is usually determined by the circuit volume, this being the number of qubits multiplied by the number of non-Clifford gates. By using additional qubits to compute chunks of logical operations that can be parallelized, these can then be applied via teleportation. The active volume is then determined by the number of logical operations making up these chunks, where each particular logical operation has an associated active volume that is measured in so-called \textit{blocks}. 

The overall cost of the computation is then dictated by the number of blocks instead of the circuit volume, which typically leads to one or more orders of magnitude reduction of the overall computation time. Note that the active volume compilation also already incorporates error correction for fault-tolerance through the surface code. The usage of this technique for runtime optimization and estimation reflects the kind of techniques that we expect could be used to compile our algorithm on a physical level, having that the specific technique and associated estimates would need to be tailored to the specific hardware under consideration. \\

The active volume compilation divides the qubits into two groups: working qubits and memory qubits. Memory qubits are used to accelerate the calculation by parallelizing blocks and storing intermediate results or ancillary states which are used in subsequent steps of the computation. The speed-up from the active volume can be estimated through the following steps:
\begin{enumerate}
    \item Identify the required number of logical qubits for implementing a circuit of interest $Q_L$. For a given budget of logical qubits $Q_B$, set the number of working qubits as $Q_W = \lfloor Q_B/2 \rfloor$. Note that the active volume compilation technique requires that $Q_W\geq Q_L$, thus at least duplicating the required number of logical qubits. The memory qubits are then allocated as $Q_M = Q_W$.
    \item Calculate the active volume in blocks $B$ for the quantum algorithm of interest. This is done by first decomposing the algorithm into elementary subroutines, which are in turn decomposed into a block-based implementation. This procedure for decomposing into blocks is covered in details in Ref.~\cite{active_volume}, with Table 1 summarizing the cost of most commonly used subroutines.
    \item Divide the number of blocks by the number of memory qubits: this will approximate the required number of cycles to implement all operations as $N_{\rm cycles}=B/Q_M$.  Note that the block-based implementation of the circuit already includes error correction through the use of a surface code, making the associated improvement unable to be converted back to elementary logical gates.
    \item Divide the number of cycles by the clock-rate $\nu_{\rm clock}$ to obtain the total runtime of the circuit $T_{\rm runtime} = N_{\rm cycles}/\nu_{\rm clock}$.
\end{enumerate}
It should be emphasized that the above constitutes a high-level description of the procedure, with additional considerations being necessary for a complete description of the active volume compilation technique. However, for the purposes of this work our description is sufficient for obtaining reasonable runtime estimates while also exploiting the associated speed-up of using the active volume compilation. 

\subsection{Time-evolution implementation} \label{subsec:time_evol}

Having described both the quantum algorithm and the qubit mapping of the vibrational degrees of freedom, the only missing piece of the workflow is the implementation of the time-evolution operator $e^{-i\hat H t}$ and the circuit for initial state preparation corresponding to $\ket{\psi_\rho}$ in \cref{eq:init_state}. As discussed in \cref{app:init_state}, this initial state can be efficiently prepared at a cost that is negligible compared with that of implementing the time-evolution. This can be understood since a small number of mean-field VSCF states can be used to represent the initial state: the initial state is extremely sparse in this basis. Given that, we now focus on the implementation of the time-evolution operator. We start by separating the Hamiltonian into its kinetic and potential components $\hat H = \hat T + \hat V$, where
\begin{align}
    \hat T &= \sum_{i\geq j} K_{ij} \hat p_i \hat p_j = \mathtt{Q_M}^\dagger \left(\sum_{i\geq j} K_{ij} \hat q_i \hat q_j\right) \mathtt{Q_M} \\
    \hat V &= \sum_{i\geq j} \Phi_{ij}^{(2)} \hat q_i\hat q_j + \sum_{i\geq j\geq k} \Phi_{ijk}^{(3)} \hat q_i\hat q_j \hat q_k \nonumber \\
    &\ \ + \sum_{i\geq j\geq k} \Phi_{ijkl}^{(4)} \hat q_i\hat q_j \hat q_k \hat q_l + \cdots,
\end{align}
where we have defined the multidimensional shifted \gls{QFT} as the product of shifted \gls{QFT}s over all modes:
\begin{equation}
    \mathtt{Q_M} = \prod_{j=1}^{M} \hat X_{s,j} \cdot \mathtt{QFT}_j.
\end{equation}
The operator $\hat V$ is already diagonal in the real-space representation, while conjugation with $\mathtt{Q_M}$ makes the kinetic term diagonal. We can then use a second-order Trotter formula to approximate the time-evolution as
\begin{equation} \label{eq:trotter_oracle}
    e^{-i\hat H t} \approx e^{-i\hat V \frac{t}{2}} e^{-i\hat T t} e^{-i\hat V \frac{t}{2}} \equiv \mhat U_2(t).
\end{equation}
In this work we chose a second-order Trotter formula since it was associated with the lowest implementation costs for achieving the required accuracy in our test systems when compared to other Trotter orders. The required evolution time for the unitary corresponding to a single time step, namely $e^{-2\pi i\hat H /\Omega}$, is not too long. This makes the required Trotter evolution time relatively short, which tends to benefit lower order Trotter formulas. We will refer to $\mhat U_2(t)$ as the Trotter oracle. 

\subsubsection{Trotter step size determination} \label{subsubsec:stepsize}
Many works have addressed the problem of bounding the error to estimate the associated Trotterization cost that achieves a target accuracy \cite{trotter_error,trotter_pt,trotter_pt_2}. However, the resulting upper bounds to the Trotter error are known to be overly pessimistic \cite{trotter_error,trotter_error_1,trotter_error_2,trotter_error_3}, with tighter estimates such as direct simulations requiring a prohibitive amount of classical resources to be obtained. One of the main contributions of this work is an alternative, cheaper-to-implement approach to more tightly control the Trotter error: instead of aiming to upper-bound the error, we estimate it using an approach based on perturbation theory. This tighter estimate is in turn used to obtain a less pessimistic step size for the Trotterized time-evolution, strongly reducing the associated cost of implementing the time-evolution operator. Inspired by the perturbation-based approach of Refs.~\cite{trotter_pt,trotter_pt_2} to estimate the Trotter error for ground-state properties, in this work we extend it to include excited states and consider a second-order Trotter formula. Still, all the discussion that follows can be straightforwardly extended to arbitrary product formulas. \\

We start by remembering that our algorithm for spectroscopy can be considered to an extent as performing phase estimation on the operator $2\pi\hat H/\Omega$. For convenience we define $\Delta t=2\pi/\Omega$, working with the operator $\hat H \Delta t$. The Trotterized time-evolution operator effectively implements a time evolution under an effective Hamiltonian, namely $e^{-i\hat H_{\rm eff}\Delta t}$. We thus need to make the eigenvalues of $\hat H_{\rm eff} $ that contribute to the absorption spectrum to be the same as those of $\hat H$ within some target accuracy $\epsilon_\nu$. To achieve this, we start by partitioning the Trotterized time-evolution into $r$ steps, from which we can write
\begin{equation}
    e^{-i\hat H_{\rm eff}\Delta t} = \left[\mhat U_2\left(\frac{\Delta t}{r}\right)\right]^r.
\end{equation}
Using the Baker-Campbell-Hausdorff formula we then arrive to
\begin{equation}
    \hat H_{\rm eff} = \hat H + \left(\frac{\Delta t}{r}\right)^2 \mhat E_2 + \mathcal{O}\left(\frac{\Delta t^4}{r^4}\right).
\end{equation}
Here we have defined the error operator
\begin{equation}
    \mhat E_2 = \frac{1}{24} \left(2[\hat T,[\hat T,\hat V]] + [\hat V,[\hat T,\hat V]]\right);
\end{equation}
different fragmentation techniques for the Hamiltonian would yield different error operators depending also on nested commutators. In the above, we adopt the straightforward fragmentation into kinetic and potential energy fragments shown in \cref{eq:trotter_oracle}. We can now use perturbation theory to estimate the eigenvalues of $\hat H_{\rm eff}$ \cite{trotter_pt, trotter_pt_2}. The explicit dependence on $\Delta t/r$ will make different degrees of perturbation theory be associated with different powers of $\Delta t/r$. First order perturbation theory is enough in this case to obtain the leading order. We thus obtain the eigenvalues of $\hat H_{\rm eff}$
\begin{equation}
    E_j^{(\textrm{eff})} = E_j + \frac{\Delta t^2}{r^2} \bra{E_j}\mhat E_2 \ket{E_j} + \mathcal{O}\left(\frac{\Delta t^4}{r^4}\right),
\end{equation}
which deviate from the true eigenvalues depending on the chosen step size $r$. For simplicity we now define the time step size $\tau_j=\Delta t/r$ such that the associated error is within some threshold, namely $|E_j^{(\textrm{eff})} - E_j| < \epsilon_\nu$. Note that the $\Delta t/r$ factor is not always smaller than one for the chosen $\Delta t$ and $r$'s in this work, meaning in principle we would need to consider higher-order contributions. However, choosing $r$ as to make the leading order error smaller than $\epsilon_\nu$ causes the remaining terms to be small as well, as long as $\epsilon_\nu$ is not too large. As shown in the simulations in \cref{sec:benchmark}, reasonable choices of $\epsilon_\nu$ yield high-quality spectra. From this, we obtain
\begin{equation} \label{eq:pec}
    \tau_j = \sqrt{\frac{\epsilon_\nu}{\bra{E_j}\mhat E_2 \ket{E_j}}}.
\end{equation}
The physical interpretation of $\tau_j$ is as the largest time step that can be taken with the Trotterized time-evolution such that eigenvalue $E_j^{({\rm eff})}$ of the effective Hamiltonian is within the allowable error threshold $\epsilon_{\nu}$ from the true eigenvalue $E_j$. Since the initial state given in \cref{eq:init_eigen} only has significant contributions from a small number of eigenstates and our final target quantity is the associated absorption spectrum, we can then choose to only account for the eigenstates making up the initial state: errors in the energies of optically inactive states will not affect the quality of the spectrum. We then choose to include each of the eigenstates with a weight associated to its overlap with the initial state $\ket{\psi_{\rho}}$ in \cref{eq:init_state}, namely
\begin{equation} \label{eq:avg_trotter}
    \tau = \sum_j |\bra{E_j} \psi_\rho\rangle|^2 \tau_j,
\end{equation}
where for simplicity we here used the same $\tau$ for all three Cartesian components $\rho=x,y,z$ by averaging the associated $\tau$'s. All of these expressions are in terms of exact eigenstates, which are obviously not accessible for systems of interest. To circumvent this, we can use approximate eigenstates as done in the initial state projection procedure. Once $\tau$ has been obtained, we can recover the associated number of Trotter steps as
\begin{equation}
    r = \left\lceil\frac{\Delta t}{\tau}\right\rceil.
\end{equation}
The inclusion of the overlap factor in \cref{eq:avg_trotter} makes it unnecessary to calculate the expression in \cref{eq:pec} for all $j$'s: it only needs to be evaluated whenever there is a significant overlap $|\bra{E_j}\psi_\rho\rangle|^2$. The expectation values corresponding to $\bra{E_j}\mhat E_2 \ket{E_j}$ can be efficiently approximated by using states obtained from \gls{VSCF} and representing the $\hat T$ and $\hat V$ operators using bosonic ladder operators. The expectation value computed in this way might differ from one expressed in the real-space basis, because the commutator algebra of the discrete operators $\hat q_j, \hat p_j$ agrees with that of the infinite-dimensional bosonic operators $b_j, b_j^\dagger$ only on a finite low-energy subspace \cite{macridin_1,macridin_2}. However, as long as enough points were used to discretize the vibrational modes when building the real-space representation, this low-energy subspace of agreement encompasses the optically active region of interest. This condition was already met when discretizing the real-space grid.

\subsubsection{Implementing time-evolution of a fragment} \label{subsubsec:fragments}

\begin{figure*}
    \centering
    \begin{quantikz}[row sep=0.4cm]
		\lstick{$\ket{q_i}$} & \qwbundle{N} & \gate[3]{\hspace{0.5cm}\mathtt{Mult}\hspace{0.5cm}} \gateinput{$a$} \gateoutput{$a$} &&&& \gate[3]{\hspace{0.5cm}\mathtt{Mult}^\dagger\hspace{0.5cm}} \gateinput{$a$} \gateoutput{$a$} & \rstick{$\ket{q_i}$} \\ 
		\lstick{$\ket{q_j}$} & \qwbundle{N} & \gateinput{$b$} \gateoutput{$b$} &&&& \gateinput{$b$} \gateoutput{$b$} & \rstick{$\ket{q_j}$} \\
		\lstick{$\ket{\vec 0}$} & \qwbundle{2N} & \gateinput{$0$} \gateoutput{$ab$} & \gate[3]{\hspace{0.5cm}\mathtt{Mult}\hspace{0.5cm}}\gateinput{$a$} \gateoutput{$a$} &  & \gate[3]{\hspace{0.5cm}\mathtt{Mult}^\dagger\hspace{0.5cm}}\gateinput{$a$} \gateoutput{$a$} & \gateinput{$ab$} \gateoutput{$0$} & \rstick{$\ket{\vec 0}$} \\
		\lstick{$\ket{\vec 0}$} & \qwbundle{b_k} & \gate{\vec X(\Phi_{ij})} & \gateinput{$b$} \gateoutput{$b$} && \gateinput{$b$} \gateoutput{$b$} & \gate{\vec X(\Phi_{ij})} & \rstick{$\ket{\vec 0}$} \\
		\lstick{$\ket{\vec 0}$} & \qwbundle{2N+b_k} && \gateinput{$0$} \gateoutput{$ab$} & \gate[2]{\hspace{0.5cm}\mathtt{Add}\hspace{0.5cm}} \gateinput{$a$} \gateoutput{$a$} & \gateinput{$ab$} \gateoutput{$0$} && \rstick{$\ket{\vec 0}$} \\
		\lstick{$\ket{R}$} & \qwbundle{b_r} &&& \gateinput{$a$} \gateoutput{$a\oplus b$} &&& \rstick{$e^{-i\Phi_{ij} q_i q_j}\ket{R}$}
	\end{quantikz}
    \caption{Quantum-arithmetic-based circuit to implement the evolution associated with an exponential $e^{-i\Phi_{ij} \hat q_i \hat q_j}$. The $
    \mathtt{Mult}$ operation corresponds to an out-of-place signed multiplier \cite{quantum_arithmetic}, and $\mathtt{Add}$ to the addition circuit for the phase gradient technique described in Appendix A of Ref.~\cite{phase_gradient}. The $\vec X(\Phi_{ij})$ operation is purely composed of Pauli X gates, directly loading the binary representation of $\Phi_{ij}$ on the quantum register using $b_k$ bits of precision. Note how the size of the registers carrying the multiplication results increases as a function of the degree of the carried monomial term.}
    \label{fig:poly_evolution}
\end{figure*}
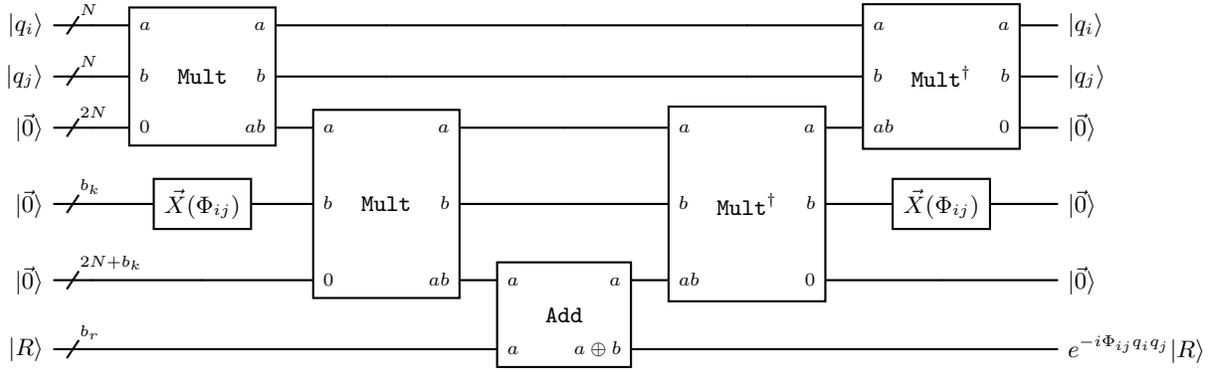

Next, we describe how to implement the exponential of a polynomial of $\hat q_j$ operators. This allows to implement all terms in the Trotter formula with the use of $\mathtt{Q_M}$ to rotate between position and momentum bases. We adapted the vibronic dynamics algorithm from Ref.~\cite{vibronic} for implementing the time-evolution of the vibrational Hamiltonian. This makes use of the efficient quantum-arithmetic-based implementation in Refs.~\cite{gidney2018halving,su2021fault}, specifically for out-of-place addition and out-of-place multiplication, as well as the phase gradient trick \cite{phase_gradient}. The basic idea of the latter is to use a resource state
\begin{equation} \label{eq:resource}
    \ket{R} \propto \sum_{j=0}^{2^{b_r}-1} e^{ij\phi_{b_r}} \ket{j},
\end{equation}
which is defined over an $b_r$-qubit register with $\phi_{b_r} = 2\pi\ell/2^{b_r}$. Here $\ell$ is an integer that carries the scale information, namely to what real numbers the integers $n$ that will be added to this state correspond to. It can be easily shown that using a modular quantum addition circuit \cite{phase_gradient} to add an integer $n$ to this state then yields the phase
\begin{equation}
    \mathtt{Add}\left\{\ket{n}\ket{R}\right\} = e^{-in\phi_{b_r}} \ket{n}\ket{R} .
\end{equation}
For simplicity we denote the state $\ket{q_j}$ as a quantum register encoding a vibrational wavefunction over the $j$-th vibrational mode, as shown in \cref{fig:grid}. Thus, by generating a register that encodes a particular polynomial term, e.g. $\Phi_{ij} \hat q_i\hat q_j$ through the use of quantum multiplication circuits, the associated phase can be implemented by using a quantum addition circuit as 
\begin{equation}
    \mathtt{Add}\left\{\ket{\Phi_{ij} q_i q_j} \ket{R}\right\} = e^{-i\Phi_{ij} q_i q_j} \ket{\Phi_{ij} q_i q_j}\ket{R}.
\end{equation}
A circuit for implementing this operation is presented in \cref{fig:poly_evolution}. The advantage of using this implementation for compiling the exponentials comes from the fact that quantum arithmetic routines have a cost scaling linearly with the number of qubits. For a given $D$-th order Taylor expansion of the PES, this approach will need $D-1$ additional registers to carry the intermediate coefficients, each requiring $N\cdot d$ qubits for $d=2,\cdots,D$. Two more registers are then needed: one for encoding the monomial coefficient $\Phi_{ij}$, alongside another register to carry the result of multiplying this coefficient by the multiplication of $q$'s. Finally, a resource state is required for each different Taylor expansion degree: the $\phi_{b_r}$ constant in each must be adjusted as to account for the $\Delta^d$ factors associated with a $d$-th degree monomial of $\hat q_j$'s [\cref{eq:qn}] such that phase gradient associated with the integer encoding $\Phi_{ij}q_iq_j$ corresponds to $e^{i\Phi_{ij}q_iq_j}$. This then entails $D-1$ resource state registers, each having $b_r$-qubits. See \cref{app:scaling} for a discussion the associated complexity of implementing these circuits.\\

In addition, intermediate coefficients can be used for implementing several different coefficients. For example, a register carrying $\ket{q_iq_j}$ can be used for implementing both terms with $\hat q_i \hat q_j$ and $\hat q_i^2 \hat q_j$. Thus, several coefficients can be implemented before passing to the uncomputation step that corresponds to the operations to the right of the addition gate in \cref{fig:poly_evolution}. We refer to this as caching of coefficients: on top of the high efficiency of the quantum arithmetic implementations, caching is an optimization that allows for an additional 50-70\% reduction in the number of independent multiplications required to implement all the terms within a single Trotter step, depending on the particular structure of the Hamiltonian coefficients. Note that the number of additions cannot be reduced through caching, as each non-zero coefficient in the Taylor representation of the Hamiltonian in \cref{eq:taylor_ham} will require one addition. A more detailed discussion on our use of caching is presented in \cref{app:caching}.
    
\section{Benchmarking the algorithm}
\label{sec:benchmark}

\subsection{Proof-of-concept simulations} \label{subsec:simulations}
\begin{figure*}[t!]
    \centering
    \includegraphics[width=0.49\linewidth]{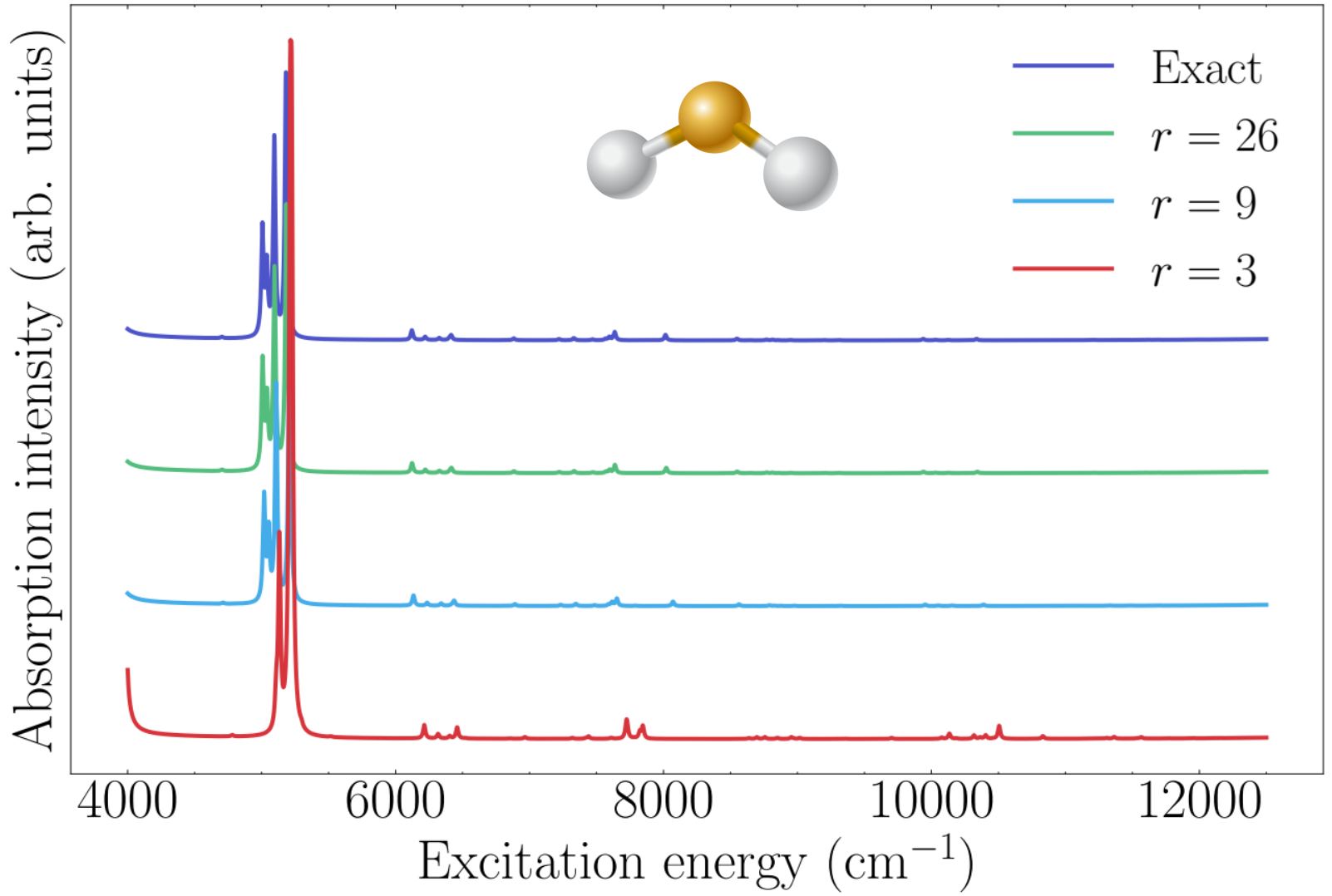}
    \includegraphics[width=0.49\linewidth]{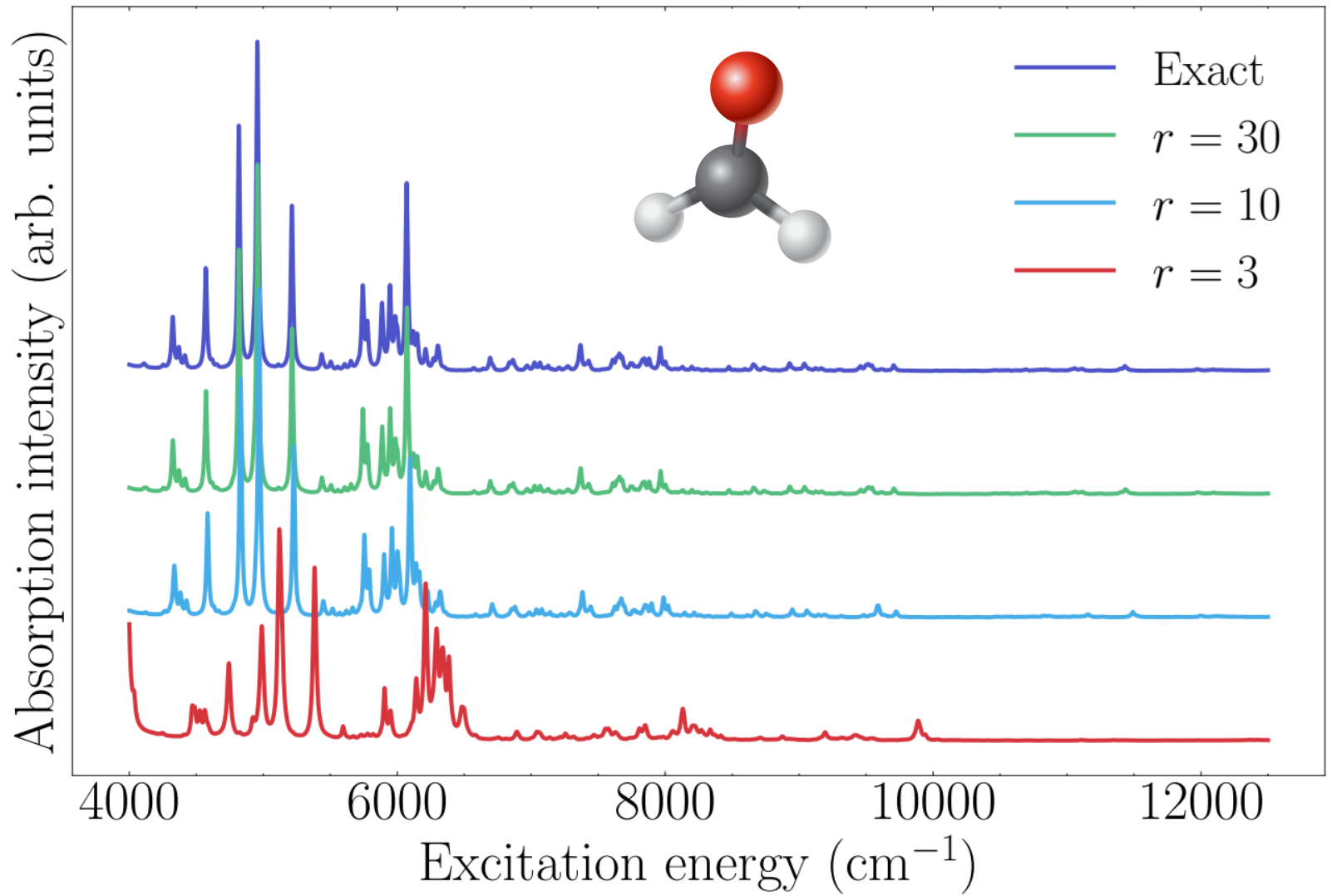}
    \caption{Simulated NIR spectra for hydrogen sulfide H$_2$S (left) and formaldehyde CH$_2$O (right) molecules using three qubits per vibrational coordinate and a fourth-degree $2$-mode Taylor expanded Hamiltonian for both. Shown $r$'s in blue, red and green were obtained using the perturbative approach in \cref{subsubsec:stepsize} for $\epsilon_\nu=1\textrm{ cm}^{-1}$, $\epsilon_\nu=10\textrm{ cm}^{-1}$ and $\epsilon_\nu=100\textrm{ cm}^{-1}$  respectively. Exact expectation values were considered when reconstructing the spectra, which amounts to taking an infinite amount of samples from the Hadamard tests. Only the NIR region has been plotted. \\}
    \label{fig:trotter}
    \includegraphics[width=0.49\linewidth]{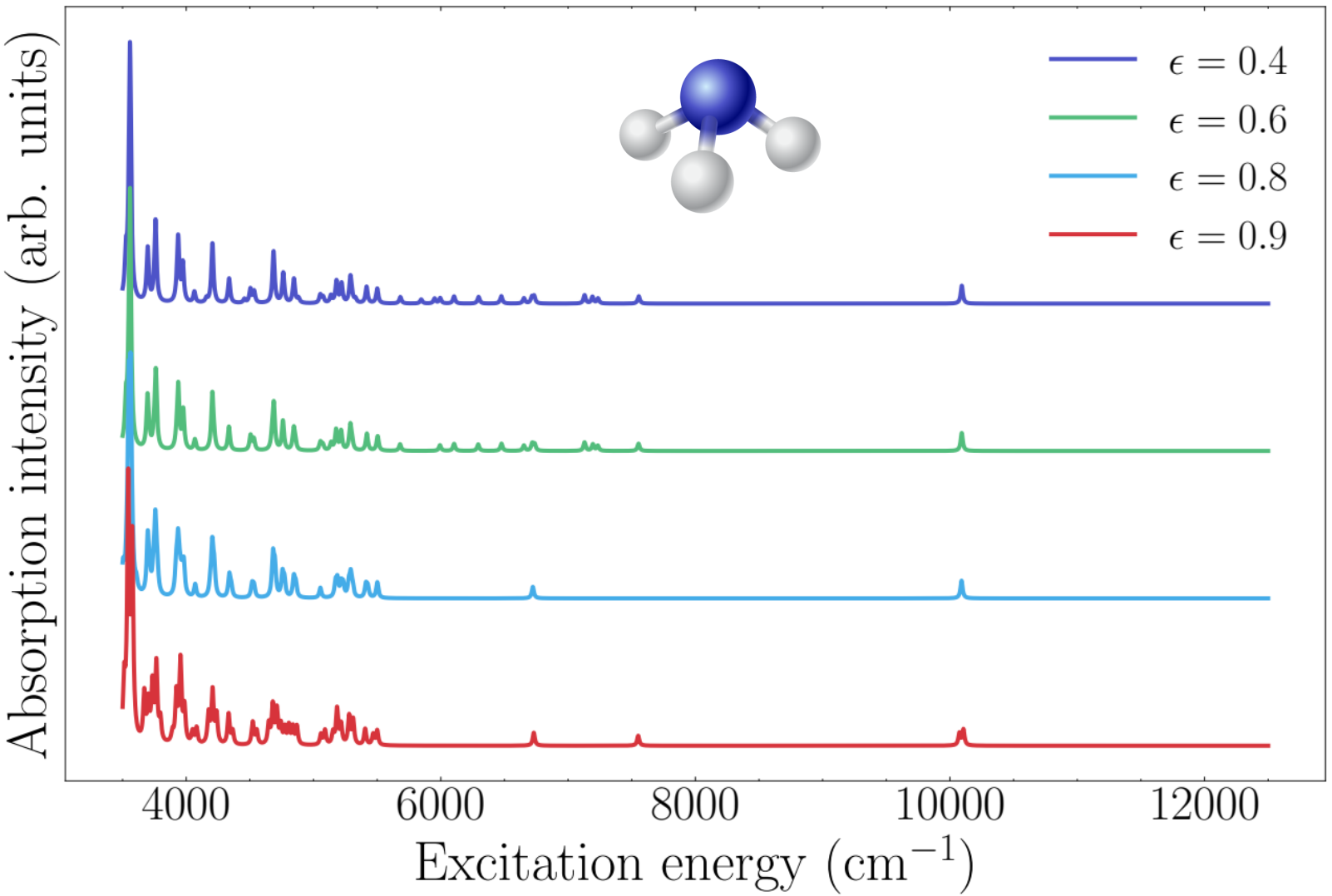}
    \includegraphics[width=0.49\linewidth]{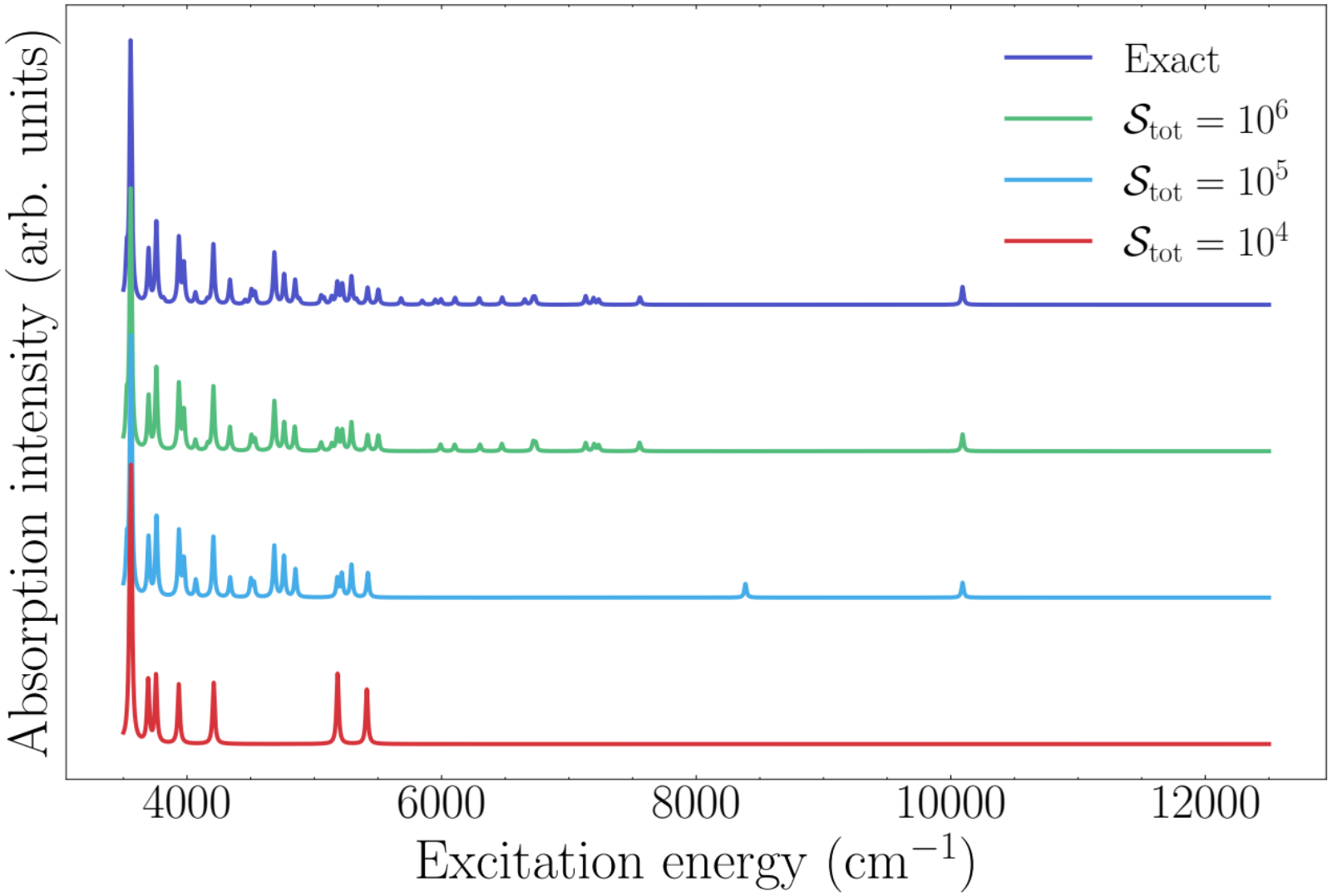}
    \caption{Reconstructed spectra in the $3'500-12'500\ {\rm cm}^{-1}$ frequency range for ammonia NH$_3$, with associated support $\Omega=9,000$ cm$^{-1}$. Explored hyperparameters consist of the Fourier integral accuracy $\epsilon$ (left) and number of shots $\mathcal{S}_{\rm tot}$ (right). Obtained spectra were post-processed with the matching pursuit algorithm from Ref.~\cite{qpe_qmegs} using a Lorentzian line shape. Only peaks with amplitudes higher than the average noise coming from finite sampling were included. \label{fig:hyper}}
\end{figure*}

In this section, we benchmark the proposed quantum algorithm by simulating the complete workflow. PennyLane \cite{pennylane} was used for simulating the quantum algorithm, obtaining vibrational Hamiltonians, and obtaining the VSCF solutions. All details for the simulations can be found in \cref{app:details}. \\

Figure \ref{fig:trotter} shows the spectra obtained by simulating the quantum algorithm for both hydrogen sulfide H$_2$S and formaldehyde CH$_2$O molecules for different values of the Trotter step size $r$. Decreasing values for $r$ are shown in these plots: these were obtained using the perturbative approach discussed in \cref{subsubsec:stepsize} for an error of respectively $\epsilon_\nu=1$ cm$^{-1}$, $\epsilon_\nu=10$ cm$^{-1}$, and $\epsilon_\nu=100$ cm$^{-1}$. For benchmarking purposes, we also include the reference simulation results (black), obtained by exact diagonalization for the smaller H$_2$S molecule where it is available, and large values of $r$ with a converged spectrum for CH$_2$O where exact diagonalization was not possible. In particular, we can see how for sensible choices of the Trotter accuracy , namely $\epsilon_\nu\leq 10\ {\rm cm}^{-1}$, the obtained spectrum is virtually indistinguishable from the exact one. This shows how $\epsilon_\nu$ can be adapted depending on the desired accuracy of the simulated spectrum, with the less stringent $\epsilon_\nu=10$ cm$^{-1}$ already yielding high-quality spectra. As expected, values of $\epsilon_\nu$ that are too large (e.g. $100\ {\rm cm}^{-1}$) bring us out of the perturbative regime, yielding low-quality spectra. These results establish an exciting proof-of-concept for the perturbation-theory-based choice of the Trotter step $r$, while also showing that our proposed workflow can indeed recover high-quality NIR spectra as desired. \\

We now discuss the selection of the hyperparameters that go into our workflow, namely the discrete Fourier transform accuracy $\epsilon$ appearing in \cref{eq:max_time} and the total number of shots $\mathcal{S}$ in \cref{eq:shot_normalization}. We here define $\mathcal{S}_{\rm tot} = 3\mathcal{S}$ to incorporate the total number of shots across all contributions to the absorption spectrum coming from the different Cartesian components of the dipole. The choice of the hyperparameters is heuristic and depends on the level of precision required, which means that for different applications these could change. However, this is purely a signal processing problem. Both of these quantities are associated to the required accuracy of the Fourier representation for a given function with some target Lorentzian broadening and spectral range. In this work these were considered to be $\eta=10\textrm{ cm}^{-1}$ and $\Omega=9,000\textrm{ cm}^{-1}$ respectively, reflecting typical values observed in NIRS. The NIR spectrum for ammonia ${\rm NH_3}$ was used for determining the hyperparameters $\epsilon$ and $\mathcal{S}_{\rm tot}$ in \cref{fig:hyper}. The hyperparameters were varied as to obtain what was deemed a high-quality replica of the exact spectrum. From this heuristic analysis, we determined that a value of $\epsilon=0.8$ and of $\mathcal{S}_{\rm tot}=10^5$ was enough to recover high-quality spectra. The algorithm from Ref.~\cite{qpe_qmegs} was used for post-processing the spectra, which obtains peak locations and amplitudes using a greedy matching pursuit algorithm with the Lorentzian line shapes.

Note that, in this work, the initial state projection technique was implemented using VSCF states that were obtained from the ground-state mean-field, as opposed to more advanced implementations of VSCF that consider a different mean-field state for targeting each excited state \cite{vscf_1,vscf_2,vscf_3,vscf_4,beyond_vscf}. This made the filtering procedure less effective, as we can see from the high amplitude in the excitation around $3'500\ {\rm cm^{-1}}$ in \cref{fig:hyper}. However, for the purposes of determining hyperparameter values, it is beyond the scope of this work to implement higher-quality approximate solutions for the vibrational Hamiltonian: our results are sufficient for showing the type of convergence that can be expected for a general NIR spectrum with Lorentzian broadening $\eta=10\ {\rm cm^{-1}}$. Finally, we note that the usage of efficient short-time Fourier transform schemes could be beneficial for lowering the required number of circuit repetitions \cite{esprit,qpe_cs}, and exploration of this avenue is left as future work. 

\subsection{Resource estimates}
\begin{table*}[]
    \centering
    \begin{tabular}{|c|c|c|c|c|} \hline
    \textbf{Molecule} & \textbf{Vibrational modes} & \textbf{Trotter $r$}& \begin{tabular}{@{}c@{}}\textbf{Maximum T-gates} \\ \textbf{(runtimes in seconds)}\end{tabular}& \begin{tabular}{@{}c@{}}\textbf{Total T-gates} \\ \textbf{(runtimes in hours)} \end{tabular}\\ \hline \hline
    H$_2$S & $3$ & $9$ &  $7.57\times10^7\ (3)$ & $1.46\times10^{12}\ (16)$\\ \hline
    NO$_2$ & $3$ & $3$ &  $2.66\times 10^7\ (1)$  & $6.02\times10^{11}\ (7)$\\ \hline
    H$_2$O & $3$ & $14$ &  $1.18\times10^8\ (5)$ & $2.19\times10^{12}\ (24)$\\ \hline    
    CO$_2$ & $4$ & $8$ &  $5.07\times10^7\ (2)$ & $9.81\times10^{11}\ (11)$\\ \hline
    CH$_2$O & $6$ & $10$ &  $2.77\times10^8\ (11)$ & $5.27\times10^{12}\ (59)$\\ \hline
    C$_2$N$_2$ & $6$ & $8$ &  $1.53\times10^8\ (6)$ & $2.97\times10^{12}\ (33)$\\ \hline
    CH$_4$ & $9$ & $13$ &  $9.97\times10^8\ (40)$ & $1.86\times10^{13}\ (207)$\\ \hline
    C$_2$H$_4$ & $12$ & $13$ &  $1.31\times10^9\ (52)$ & $2.45\times10^{13}\ (272)$\\ \hline
    HC$_2$N$_3$ & $12$ & $6$&  $8.47\times10^8\ (34)$ & $1.70\times10^{13}\ (189)$\\ \hline
    \end{tabular}
    \caption{Resource estimates for optimized algorithm presented in this work. The maximum T-gates column represents the number of T-gates associated with the deepest circuit in the workflow, i.e. the one coming from the maximum time $k_{\rm max}$, with total T-gates considering all the circuit repetitions for reconstructing the spectrum. Associated runtimes are shown in parenthesis, which were calculated using the active volume compilation technique from Ref.~\cite{active_volume} discussed in \cref{subsubsec:act_vol} considering a budget of $500$ logical qubits. For concreteness we assumed a logical clock-rate of $\nu_{\rm clock}=1$ MHz for the hardware, noting that runtimes are inversely proportional to this quantity. A basis of 4 qubits was used for discretizing each vibrational mode as shown in \cref{subsec:ham}, alongside a value of $\epsilon_\nu=10$ cm$^{-1}$ for the perturbative of the Trotter step discussed in \cref{subsubsec:stepsize} and a total of $\mathcal{S}_{\rm tot}=10^5$ shots distributed across all different times and Cartesian components of the dipole.}
    \label{tab:estimates}
\end{table*}

For ease of calculation of our resource estimates, in this work we first decomposed the circuit into T-gates. We then considered an average usage of 10 blocks per T-gate following the active volume architecture in Ref.~\cite{active_volume}. Note that a more high-fidelity estimation needs to distinguish the active volume of each separate subroutine; efforts to this effect are a necessary next step for further strengthening the resource estimates but is outside the scope of the present work. Still, we expect the numbers obtained here to be highly representative and within the same order of magnitude of what would be obtained with a high-fidelity full-stack resource estimation. \\

So far we have described a highly optimized algorithm for obtaining NIR spectra on a quantum computer. \cref{tab:estimates} shows the associated resource estimates for a variety of molecules. Overall, under the $1$ MHz logical clock-rate assumption, our presented runtimes are of the order of a few seconds/minutes for single circuits and of a few hours/days for the entire workflow. When also considering the circuit requirements of few hundreds of logical qubits and few billions of T-gates per circuit, this establishes the simulation of NIR spectra as a promising application of early fault-tolerant quantum computers for going beyond current classical computing capabilities.

    \glsresetall
	\section{Conclusions}
    \label{sec:conclusions}

Simulations of vibrational spectra of industrially relevant molecules could be a cheap, safe and rapid source of training data for statistical chemical detection models. Here we have proposed a highly optimized quantum algorithm to simulate NIR spectra on a quantum computer.  Our algorithm has an $\mathcal{O}(M^2)$ scaling with respect to the number of vibrational modes $M$, in comparison to the $\mathcal{O}(M^{12})$ from high-accuracy classical methods that are required for this application. Moreover, the small prefactor costs in our algorithm make the overall cost to be extremely attractive for implementation on early fault-tolerant quantum computers. This low cost, together with the strong anharmonicity of the NIR presenting challenges for traditional classical methods, make simulating vibrational spectra a promising application of quantum computers in the early fault-tolerant era. \\

While qubit counts of a few hundred and T-gate counts of a few billions are a major reduction in algorithm cost, further optimizations could be performed from working at the intersection of the abstract algorithm and the physical architecture, mediated through quantum error correction. The active volume technique exploited in this paper is one example of this: we anticipate that dedicated efforts aimed at compiling the algorithm down to the physical level and targeting specific hardware can bring further optimizations, while also resulting in more accurate runtime estimates. In addition we also expect the usage of techniques that approximate Fourier transforms using a reduced number of points to be useful for lowering the overall cost of our algorithm \cite{qpe_cs,esprit,super_resolution}. \\

Overall, quantum-computing-enhanced cheap simulations of NIR spectra could significantly improve NIR-based chemical detection. Augmenting training datasets with simulated data has the potential to improve model detection accuracy, as well as improve the model's transferability to new spectrometer devices, deployment contexts, or environmental conditions. The cheap, fast, non-invasive nature and wide applicability of NIR spectroscopy also means that such quantum data augmentation could be beneficial in a wide range of applications beyond chemical detection, such as in-line monitoring of chemical concentrations to optimize reaction yield in pharmaceutical applications \cite{petersen2010situ,simon2015assessment}, detection of food product adulteration \cite{woodcock2008better,grabska2021theoretical,bec2022miniaturized}, or even more exotic use cases like fine art analysis \cite{cucci2016reflectance,polak2017hyperspectral}. This work is an early step in applying quantum computers to unlocking the full potential of NIR-based spectroscopy as a high-fidelity, broadly applicable chemical characterization technique across a variety of industries. \\

\section*{Acknowledgements}
The authors are grateful to Artur Izmaylov, Shreyas Malpathak, Sangeeth Das Kallullathil, Ewan Murphy, Evgenia Krichanovskaya, and Tarik El-Khateeb for stimulating discussions.

	\bibliography{biblio}

    \onecolumngrid
	\appendix

\section{Algorithm scaling} \label{app:scaling}
In this appendix we deduce the scaling of our algorithm with respect to the number of vibrational modes $M$, with $N$ the associated number of qubits representing each mode. We will be considering a general $n$-mode expansion of the Hamiltonian. We first start by counting the number of terms in the Taylor Hamiltonian, considering $D$ the maximum degree appearing in the Hamiltonian. We will count the number of terms that can appear for each degree $d=1,\cdots,D$. For $d=1$ there are $0$ terms, since these are related to first-order derivatives which vanish for the equilibrium geometry. For $d=2$ we then have all the possible $\hat q_i\hat q_j$ combinations for $i > j$ plus all the $\hat q_i^2$, yielding ${M\choose 2} + M =M(M+1)/2 \sim \mathcal{O}(M^2)$ terms. For $d=3$, we have all the possible combinations corresponding to the following cases: i) $\hat q_i \hat q_j \hat q_k$, ii) $\hat q_i^2 \hat q_j$, 3) $\hat q_i \hat q_j^2$, and 4) $\hat q_i^3$ for $i>j>k$. However, the first term will only appear if $n\geq 3$. The associated complexity then becomes $\mathcal{O}(M^3)$ if $n\geq 3$, and $\mathcal{O}(M^2)$ if $n=2$. Following this procedure it becomes easy to show that in general the number of terms will be given by $\mathcal{O}(M^n)$. Noting that the main cost of the algorithm will be given by the implementation of the quantum arithmetic multiplications, we now deduce the scaling of the number of multiplications appearing in our algorithm. The terms with the most multiplications in them correspond to the $D$-th degree polynomials, which require to multiply quantum registers with the following sizes: 
\begin{equation}
(N\times N)\rightarrow(2N\times N)\rightarrow\cdots\rightarrow((D-1)N\times N)\rightarrow (DN\times b_k),    
\end{equation}
where $b_k$ is the size of the register encoding the Taylor polynomial constant which in this work was selected as $b_k=10$. Noting that multiplying two registers of size $A$ and $B$ has an associated T-gate cost of $2A B-\max\{A,B\}\sim\mathcal{O}(AB)$, the complexity of this multiplication sequence is then
\begin{equation}
\mathcal{O}(N^2 + 2N^2 + \cdots + (D-1)N^2 + DNk) = \mathcal{O}(D(D-1)N^2+DNb_k).
\end{equation}
We now arrive to the scaling of implementing a Trotter step $\mathcal{O}\big((D(D-1)N^2+DNb_k)\cdot M^n\big)$, which is obtained by multiplying the number of elements with the multiplication cost of each. The scaling of the algorithm will then be given by the total required number of Trotter steps that need to be implemented. This quantity will also depend on the number of points discretizing the time grid in the discrete Fourier transform, namely $k_{\max}$ in \cref{eq:kmax}, the total number of shots $\mathcal{S}$ with its associated shot allocation shown in \cref{eq:shot_normalization}, and the Trotter step size $r$ that was obtained perturbatively for each system. Note that fixing the frequency window in the NIR region makes $k_{\max}$ and $\mathcal{S}$ not be system dependent, namely $k_{\max},\mathcal{S}\sim\mathcal{O}(1)$. Furthermore, from the results obtained in \cref{tab:improvements} it can be seen how the Trotter $r$ does not seem to increase with system size, from which we can empirically claim that $r\sim\mathcal{O}(1)$. From these considerations and noting that the register size $k$ is also a constant that is fixed, we arrive to the final complexity for our algorithm as
\begin{equation}
    C[D,N,M]\sim\mathcal{O}(D^2N^2M^n).
\end{equation}
In particular, for the $2$-mode fourth-degree Taylor Hamiltonians discussed in this work, we get a complexity with respect to the number of modes scaling as $\mathcal{O}(M^2)$. \\

Having deduced the complexity of the quantum algorihtm, we now comment on the complexity of classical VCI algorithms. We consider the VCI hierarchy of methods, having VCI-SD, VCI-SDT, and VCI-SDTQ, and so on, which incorporate up to double, triple, and quadruple excitations respectively. As discussed in \cref{sec:application}, it has been shown that at least quadruple excitations are required for providing the required accuracy. The matrix representing the Hamiltonian associated with VCI-SDTQ consists of $\mathcal{O}(M^4)$ elements, which must then be diagonalized: the NIR spectrum contains a combinatorial number of overtone and combination band excited states that are located deep in the spectrum. Considering the well known cubic scaling for diagonalization routines, the associated scaling of this method then becomes $\mathcal{O}(M^{12})$.
    
	\section{Obtaining the Taylor Hamiltonian} \label{app:taylor}
	In this appendix we show how to obtain the vibrational Hamiltonian from first principles. We start by considering the molecular Hamiltonian within the Born-Oppenheimer approximation. This means the electronic state is the ground state throughout different nuclear geometries. From this, a PES is defined, from which we can write the nuclear Hamiltonian as
	\begin{equation}
		\hat H = \hat T_N + \hat V(\mathbf{R}),
	\end{equation}
	where $\mathbf{R}=\{\vec R_1,\cdots,\vec R_N\}$ represents the set of nuclear Cartesian coordinates, $\hat V(\mathbf{R})$ is the PES for the electronic ground state, and 
	\begin{equation}
		\hat T_N=\sum_{\alpha=1}^{N_{\rm atoms}} -\frac{\hbar^2}{2M_\alpha} \nabla_\alpha^2
	\end{equation}
	is the kinetic energy operator, for $M_i$ the mass of nucleus $\alpha$. Note that despite the lack of involvement from multiple electronic states during \gls{NIR} experiments, including terms from the diagonal Born-Oppenheimer correction which go beyond the Born-Oppenheimer approximation has been shown to affect the overall quality of the PES and associated vibrational quantities \cite{dboc_1,dboc_2}. Adding these corrections entails modifying the values of the $\hat V(\mathbf R)$ term, which does not affect the remainder of our discussion. The first step for defining the vibrational Hamiltonian is to find the equilibrium geometry $\mathbf{R}^{(eq)} \equiv \{\vec R_1^{(eq)},\cdots,\vec R_N^{(eq)}\}$. Normal coordinates are found by diagonalizing the Hessian in this equilibrium geometry, obtaining the vibrational modes of the molecule along with the associated frequencies $\omega_i$ and effective masses $\mu_i$. These correspond to $M=3N_{\rm atoms}-6$ different modes ($M=3N_{\rm atoms}-5$ for linear molecules) $Q_i$'s, which are related by the transformation
	\begin{equation} \label{eq:normal_modes}
		R_{\alpha\rho} = R_{\alpha\rho}^{(eq)} + \sum_{i=1}^M B_{\alpha\rho,i} Q_i,
	\end{equation}
	where $\alpha$ is an index that runs over the different atoms, $\rho=x,y,z$ is the Cartesian component, and $B_{i,\alpha\rho}$ is an element of $3N_{\rm atoms}\times M$-dimensional matrix which associates the Cartesian displacement of nuclear coordinate $R_{\alpha\rho}$ with the normal coordinate $i$.
	Neglecting the rotational and rovibrational contributions to the Hamiltonian and defining the normal coordinates in natural units
	\begin{equation}
		q_i = \sqrt{\frac{\omega_i}{\hbar}} Q_i,
	\end{equation}
	we arrive to the Taylor-expanded form of the vibrational Hamiltonian
	\begin{align}
		\hat H &= \hat H_{\rm harm} + \hat H_{\rm anh}, \\
		\hat H_{\rm harm} &= \frac{1}{2}\sum_{i=1}^M \hbar\omega_i \left(\hat p_i^2 + \hat q_i^2\right), \\
		\hat H_{\rm anh} &= \sum_{i\leq j\leq k}
		\Phi_{ijk}^{(3)} \hat q_i\hat q_j\hat q_k + \sum_{i\leq j\leq k\leq l} \Phi_{ijkl}^{(4)} \hat q_i\hat q_j\hat q_k \hat q_l + \cdots, 
	\end{align}
	for $\hat p_i^2 = - \partial^2_{q_i}$ the associated momentum in natural units for the normal mode $q_i$. We then have a mapping where we can express $\mathbf{R} = \mathbf{R}(q_1,\cdots,q_M)$, such that $\mathbf{R}^{(eq)} = \mathbf{R}(0,\cdots,0)$.  \\
	
	Once the normal coordinates have been determined, various different procedures can be used to obtain the Taylor expansion coefficients in \cref{eq:taylor_ham}. In this work we performed a multi-dimensional polynomial fit of the $\Phi$ coefficients using an $n$-mode expansion of the PES
	\begin{equation}
		\hat V(q_1,\cdots,q_M) = \hat V_0 + \sum_{i=1}^M \hat V_1^{(i)}(q_i) + \sum_{i>j} \hat V_2^{(i,j)}(q_i,q_j) + \sum_{i<j<k} \hat V_3^{(i,j,k)}(q_i,q_j,q_k) + \cdots,
	\end{equation}
	which has been shown to recover the properties from the full multi-dimensional PES with a low expansion order \cite{vmp2}. Here we defined each of the terms in the expansion as
	\begin{align} \label{eq:0-mode}
		\hat V_0 &\equiv \hat V(q_1=0,\cdots,q_M=0) = \hat V(\mathbf{R}^{(eq)}) \\
		\hat V_1^{(i)}(q_i) &\equiv \hat V(0,\cdots,0,q_i,0,\cdots,0) - \hat V_0 \\
		\hat V_2^{(i,j)}(q_i,q_j) &\equiv \hat V(0,\cdots,q_i,\cdots,q_j,\cdots,0) - \hat V_1^{(i)}(q_i) - \hat V_1^{(j)}(q_j) - \hat V_0  \label{eq:2_mode} \\
		\nonumber \vdots \, .
	\end{align}
	Finally, we note that the same $n$-mode expansion can be used to arrive to a Taylor form for the dipole operator over each of its Cartesian components.

\section{Mode localization} \label{app:mode_loc}
The mode localization technique \cite{mode_loc_1,mode_loc_2,mode_loc_3,mode_loc_4} rotates the vibrational modes making the Hamiltonian more sparse. It has been shown that the Hamiltonian obtained using the mode localization technique has similar accuracy to considering normal modes while going to a higher degree in the $n$-mode expansion in Eqs.(\ref{eq:0-mode}-\ref{eq:2_mode}) \cite{mode_loc_1}, which effectively allows us to work with a lower expansion degree while retaining accuracy. This technique works in perfect analogy with orbital localization in electronic structure \cite{foster_boys,pipek_mezey}: we go from (unlocalized) molecular orbitals that are spread throughout the entire molecule, to localized orbitals which will only interact strongly with nearby orbitals. In this work we thus make use of the mode localization technique, which defines the local coordinates $\tilde q_i$ through a unitary transformation as
	\begin{equation}
		\tilde q_i = \sum_{j=1}^M U_{ij} \hat q_j 
	\end{equation}
	The localizing unitary is found by maximizing the sum of atomic contributions to the modes in analogy to the Pipek-Mezey orbital localization technique in electronic structure \cite{pipek_mezey}, 
    \begin{equation}
		\mathbf{U}^* = \argmax_{\mathbf{U}} \left\{\sum_{j=1}^M  \sum_{\alpha=1}^{N_{\rm atoms}} \left(\sum_{\rho=x,y,z} \left(\sum_{i=1}^M U_{ij}\tilde B_{\alpha\rho,i}\right)^2\right)^2\right\},
	\end{equation}
    where $U_{ij}$'s are the elements of the $M\times M$ matrix $\mathbf{U}$ that is being optimized, and $\tilde B_{\alpha\rho,i}$ corresponds to the displacement vectors in \cref{eq:normal_modes} that have been normalized as
\begin{equation}
    \tilde B_{\alpha\rho,i} = \frac{B_{\alpha\rho,i}}{\sqrt{\sum_{\tilde\alpha\tilde\rho} |B_{\tilde\alpha\tilde\rho,i}|^2}}.  
\end{equation}
    Performing this transformation means that the Hessian is no longer diagonal, which means terms of the form $\tilde p_j\tilde p_k$ for $j\neq k$ now appear in the kinetic energy operator. The Hamiltonian can now be written as
	\begin{align} 
		\hat H = \sum_{i\geq j} \left( K_{ij} \tilde p_i \tilde p_j + \tilde\Phi_{ij}^{(2)} \tilde q_i \tilde q_j \right) + \sum_{i\geq j\geq k} \tilde\Phi_{ijk}^{(3)} \tilde q_i \tilde q_j \tilde q_k + \sum_{i\geq j\geq k\geq l} \tilde\Phi_{ijkl}^{(4)} \tilde q_i \tilde q_j \tilde q_k \tilde q_l  + \cdots,
	\end{align}
	where the $\tilde\Phi$ coefficients are obtained from a Taylor expansion as in the normal coordinates case, while the kinetic energy constants corresponds to
	\begin{equation}
		K_{ij} = \sum_{k=1}^M \frac{\hbar \omega_k}{2} U_{ki} U_{kj}
	\end{equation}
for the matrix elements of the optimized matrix $\mathbf{U}^*$. The mode localization procedure effectively yields a Hamiltonian with the same structure as in \cref{eq:taylor_ham} where normal coordinates were used, with the only differences being the value of the coefficients, and the kinetic coefficients $K_{ij}$'s not being diagonal anymore. As such, the treatment of these two different representations is completely equivalent from an algorithmic perspective and normal modes can be generally replaced by localized modes without any additional considerations.

\section{Comparing Hamiltonian representations}
\label{app:cf_reps}

There are three main ways of representing the vibrational Hamiltonian. Starting from the ground-state PES $E(q_1,\cdots,q_M)$, expressed in terms of normal mode coordinates $q_i$, the conceptually simplest is to directly represent the Hamiltonian in terms of position and momentum operators $\hat q_i, \hat p_i$ -- this was described in \cref{app:taylor}. As mentioned in the main text, the other canonical representation is obtained by casting the Taylor form in terms of the bosonic ladder operators 
\begin{equation}
    \hat q_i = \frac{1}{\sqrt{2}}(b_i^\dagger + b_i), \quad
    \hat p_i = \frac{i}{\sqrt{2}}(b_i^\dagger - b_i).
\end{equation}
With this, the Hamiltonian is given by
\begin{equation}
    \hat H = \frac{1}{2}\sum_{i=1}^M \hbar\omega_i b_i^\dagger b_i + \sum_{i\leq j\leq k}
    \Phi_{ijk}^{(3)} (b_i^\dagger + b_i)(b_j^\dagger + b_j)(b_k^\dagger + b_k) + \sum_{i\leq j\leq k\leq l} \Phi_{ijkl}^{(4)} (b_i^\dagger + b_i)(b_j^\dagger + b_j)(b_k^\dagger + b_k)(b_l^\dagger + b_l) + \cdots.
\end{equation}
where for brevity the $\Phi$ coefficients absorb the square root factors from the operator transformation. 

The third Hamiltonian form is the so-called Christiansen form, described briefly in the main text and introduced in Ref.~\cite{christiansen}. We now considering the maximal number of modals $N_{\text{max}}$ for each mode $m$ with associated modals $\phi_{m,l}(q_m)$, where we then have $l=1,\cdots,N_{\rm max}$. A register of size $N_{\text{max}}$ is then associated to each mode to represent which of the modals is occupied. For example, an occupation vector $\ket{n_m} \equiv \ket{0,0,\dots,0,1,0,\dots,0}$ for the $m$-th mode with a $1$ in the $l$-th position means that the $m$-th mode is in its $l$-th modal. Occupation number vectors for each mode are then concatenated together to give the full state vector $\ket{\mathbf{n}} = \ket{n_{1,l}, n_{2,l}, ...}$. Together with the occupation number vectors, we define a set of bosonic-like creation and annihilation operators $a_{m,l}^\dagger, a_{m,l}$ that modify these modal occupation numbers, 
\begin{equation}
    a_{m,l}^{\dagger}\ket{\mathbf{n}} = \ket{\mathbf{n} + \mathbf{1}_{m,l}} \delta_{0, \{m,l\}} \quad a_{m,l}\ket{\mathbf{n}} = \ket{\mathbf{n} - \mathbf{1}_{m,l}} \delta_{1, \{m,l\}}
\end{equation}
where $\ket{\mathbf{1}_{m,l}}$ is a full state vector with all zeros except in the $l$-th position of the $m$-th mode register. In this representation we recognize that since, on physical grounds, each mode can only be in a single modal at once, each mode register $\ket{n_m}$ can host only a single $1$. This, in effect, imposes a Pauli-like exclusion principle, together with the maximum mode occupancy of $1$ per mode register. The main advantage of this approach is that the representation makes the operator algebra very similar to the electronic structure case (with the added simplification of commuting operators), allowing a wholesale import of computational techniques such as configuration interaction, coupled cluster, and even vibrational density-matrix renormalization group. 

Employing the Christiansen representation, the Hamiltonian takes the form
\begin{align}
    \hat{H} &= \sum_{m}^M \sum_{l,l
'}^{N_{\text{max}}} H_{m,l,l'} a_{m,l}^\dagger a_{m,l} + \sum_{m<n}^M \sum_{l,l
',k,k'}^{N_{\text{max}}} H_{m,n,l,l',k,k'} a_{m,l}^\dagger a_{n,k}^\dagger a_{m,l'} a_{n,k'} \nonumber \\
&\ \ + \sum_{m<n<p}^M \sum_{l,l
',k,k',j,j'}^{N_{\text{max}}} H_{m,n,p,l,l',k,k',j,j'} a_{m,l}^\dagger a_{n,k}^\dagger a_{p,j}^\dagger a_{m,l'} a_{n,k'} a_{p,j'} + \cdots
\end{align}
where 
\begin{align}
    H_{m,l,l'} =& \int_{-\infty}^{\infty} \, dq_m \, \phi_{m,l}(q_m) \left( \hat T(q_m) + \hat V_1(q_m) \right) \phi_{m,l'}(q_m), \\
    H_{m,n,l,l',k,k'} =& \int_{-\infty}^{\infty} \int_{-\infty}^{\infty} \, dq_n dq_m \, \phi_{m,l}(q_m) \phi_{n,k}(q_n)  \hat V_2(q_m, q_n)  \phi_{m,l'}(q_m) \phi_{n,k'}(q_n), \\
    H_{m,n,p,l,l',k,k',j,j'} =& \int_{-\infty}^{\infty} \int_{-\infty}^{\infty} \int_{-\infty}^\infty \, dq_n dq_m dq_p \, \phi_{m,l}(q_m) \phi_{n,k}(q_n) \phi_{p,j}(q_p)  \hat V_3(q_m, q_n, q_p) \phi_{m,l'}(q_m) \phi_{n,k'}(q_n) \phi_{p,j'}(q_p), \\
    \nonumber \vdots \ ,
\end{align}
where $\hat T$ is the kinetic energy operator and $\hat V_{n}$'s are the components obtained from expanding the potential energy in the $n$-mode framework as shown in Eqs.(\ref{eq:0-mode}-\ref{eq:2_mode}). While the potential energy terms may be those of the Taylor approximation, they do not have to be: thus the number of terms in the Hamiltonian does not depend on the accuracy of the potential energy representation, and is always scaling as $M^n N_{\text{max}}^{2n}$. This could be constructed as an advantage of the Christiansen representation, in that it can readily make use of a highly accurate PES. Still, it is highly useful to have a trade-off of accuracy versus cost, which is exactly what both the Taylor form and the bosonic form allow by including Taylor expansions of increasing degree to approximate the PES with increasing accuracy at the cost of more terms in the Hamiltonian. 

The explicit form of the Hamiltonians in the different representations makes their advantages and disadvantages readily apparent. Neither of the second-quantized forms of the Hamiltonian are amenable to a direct quantum arithmetic circuit implementation of the time-evolution: instead, it is necessary to pick a truncation $N_{\text{max}}$ for the maximal allowed number of modals, employ some boson-to-qubit conversion scheme \cite{sawaya2020resource,ibm_vibrational} and then resolve the time-evolution as a series of Pauli rotations. The number of such rotations scales directly with the number of terms in the Hamiltonian, which for an $n$-mode Hamiltonian is asymptotically $M^n N_{\text{max}}^{2n}$. There are fragmentation-like techniques that allow a reduction of the number of terms \cite{artur_vib}, much like double factorization and related approaches in electronic structure, however even in that case the cost turns out to be severe -- and to require very specific kinds of unitary operations.

By contrast with both of these, the grid discretization of each mode coordinate in the Taylor form means that the matrix elements of the Hamiltonian with respect to individual grid points are encoded directly in the eigenvalues of the position and momentum operators, rather than being made explicit as individual Pauli rotations. This means that the effective number of individual arithmetic operators required to implement all the potential energy terms of the same Hamiltonian scales as $M^n \left(\log N_{\text{grid}}\right)^{2n}$. This does come at the cost of a more complicated initial state preparation routine, as the amplitudes at each grid point assumed by the given modal need to be encoded into the initial state wavefunction. However, this still occurs on a per-mode-register basis, rather than in the entire Hilbert space (see more details about initial state preparation in \cref{app:init_state}). Since the number of qubits required to accurately capture a sufficiently large number of modals scales logarithmically with the number of modals, $N_{\text{grid}} \sim \log N_{\text{max}}$, this cost is usually negligible compared with the cost of implementing the time-evolution operator.

\section{Additional details for quantum algorithm}

\subsection{Maximum evolution time} \label{app:max_time}
We now show how the maximum evolution time in \cref{eq:max_time} was chosen. We start by writing the condition for the time truncation error to be smaller than some target $\epsilon$:
\begin{equation}
    \left | \int_{-\infty}^{\infty} dt e^{i\omega t} \frac{e^{-\eta |t|}}{2} \bra{\psi}e^{-i\hat H t} \ket{\psi} - \int_{-T_{\rm max}}^{T_{\rm max}} dt e^{i\omega t} \frac{e^{-\eta |t|}}{2} \bra{\psi}e^{-i\hat H t} \ket{\psi} \right | \leq \epsilon.
\end{equation}
We can then combine the limits of the integrals and apply the Cauchy-Schwarz inequality to arrive to
\begin{equation}
    \int_{T_{\rm max}}^\infty dt \left|e^{-\eta t} \right| \leq \epsilon,
\end{equation}
where we have used the fact that $|e^{i\omega t}| = 1$ and that $|\bra{\psi}e^{-i\hat H t}\ket{\psi} \leq 1$. Performing this integral, we arrive to
\begin{equation}
    \frac{e^{-\eta T_{\rm max}}}{\eta} \leq \epsilon,
\end{equation}
from which we can arrive to the condition
\begin{equation}
    T_{\rm max} \geq \frac{1}{\eta} \log \frac{1}{\eta\epsilon}.
\end{equation}
Note that this integral will have inverse energy units. This becomes apparent if we consider the case where we have a single peak, which would be obtained when considering any eigenstate $\ket{\psi}=\ket{E_k}$. In this case, the infinite integral becomes
\begin{equation}
    \int_{-\infty}^{\infty} dt e^{i\omega t} \frac{e^{-\eta |t|}}{2} \bra{E_k}e^{-i\hat H t}\ket{E_k} = \frac{\eta}{\eta^2 + (\omega - E_k)^2}.
\end{equation}
This Lorentzian reaches its maximum at $\omega=E_k$ with an associated value of $1/\eta$. Rescaling $\epsilon$ by this quantity effectively makes it a quantity without dimensions, from which we can write the expression for the maximum evolution time
\begin{equation}
    T_{\rm max} = \frac{1}{\eta}\log\frac{1}{\epsilon}
\end{equation}
as stated in \cref{eq:max_time}.

\subsection{Initial state preparation} \label{app:init_state}
We now present a routine for efficiently preparing the initial state for the quantum algorithm, which corresponds to the dipole acting on the ground state as shown in \cref{eq:init_state}. We refer to this routine as the \textit{indicator state preparation} since we will use additional registers to flag on which modes different contributions of the dipole operator acts. We start by writing the dipole operator using a first-degree Taylor expansion as
\begin{equation}
    \hat\mu_\rho \approx \sum_{i=1}^M m_{i}^{(\rho)} \hat q_i,
\end{equation}
where the $m_{i}^{(\rho)}$ coefficients can be obtained using the same techniques that were used for obtaining the Taylor representation of the Hamiltonian in \cref{app:taylor}. Note that higher order expansions can be used for a better representation of the dipole. However, using a linear expansion is a common practice which already yields good results \cite{sibert2023modeling,veit2020predicting}. We thus consider just a first-degree expansion for simplicity, noting that our algorithm is straightforward to generalize to higher orders. The indicator state preparation approach exploits two facts about the initial state in \cref{eq:init_state} corresponding to the molecular dipole acting on the vibrational ground state:
 \begin{itemize}
     \item The initial state $\ket{\psi_\rho}$ consists of a sparse linear combination of product states, which can be written as
\begin{equation} \label{eq:dip_init}
    \ket{\psi_\rho} = \sum_{i=1}^M  c_{i}^{(\rho)} \ket{\phi_{i,1}}_i \bigotimes_{j\neq i} \ket{\phi_{j,0}}_j,
\end{equation}
where $i$ and $j$ run over the $M$ different vibrational modes, and the ket sub-indices indicate the qubit register associated with a given vibrational mode. The sparse structure makes this state easy to prepare by using a QROM-based circuit \cite{qrom} to load the $c_i^{(\rho)}$ coefficients on a qubit register labeling $i$ and preparing each $i$-dependent product state controlled on this register.
     \item As seen in \cref{eq:dip_init}, the product states making up $\ket{\psi_\rho}$ are extremely similar to each other, which can be exploited to further simplify the circuit.
 \end{itemize}
Note that the projection procedure for filtering out the mid-IR states outlined in \cref{subsec:qpe} should not change the structure of this state. The overall procedure for preparing this initial state then consists of the next steps: 
\begin{enumerate}
    \item Run the Vibrational Self-Consistent Field (VSCF) method to obtain the product states $\ket{\phi_{i,0}}_i$ associated to the ground-state of each vibrational mode.  
    \item Represent the molecular dipole operator using the selected vibrational modes. This yields the coefficients $c_i^{(\rho)}$ and the normalized wavefunctions $\ket{\phi_{i,1}}_i$ such that $c_i^{(\rho)}\ket{\phi_{i,1}}_i = m_i^{(\rho)}\hat q_i\ket{\phi_{i,0}}_i$.
    \item Implement a transformation on each of the $M$ registers which acts on the initialized all-zeros computational basis state as $\ket{\vec 0}_i\rightarrow \ket{\phi_{i,0}}_i$. This operation can be implemented using e.g. generic state preparation routines \cite{mottonen} or optimized variational circuits \cite{variational_prep}.
    \item Use a coefficient preparation circuit to load the coefficients $c_i^{(\rho)}$ on an ancillary quantum register as $\ket{\vec 0}_c\rightarrow \sum_{i} c_{i}^{(\rho)} \ket{i}_c$. This can be done using many techniques, such as generic state preparation, variational techniques, or QROM-based $\mathtt{PREPARE}$ circuit implementation \cite{qrom}.
    \item Initialize $M$ ``indicator'' qubits in a separate register $\ket{\vec 0}_I$, which will be used for flagging when the state $\ket{\phi_{i,1}}_i$ will be implemented. Use QROM controlled on the coefficients register $\ket{}_c$ such that $\ket{i}_c \ket{\vec 0}_I\rightarrow \ket{i}_c\ket{b_i}_I$, where $b_i$ consists of all zeros with a single $1$ in qubit $i$. This step can also be included in step $3$ during the coefficient loading step if a QROM-based technique was used.
    \item Controlled on the $i$-th qubit in the indicator register, apply a transformation which acts on the $i$-th vibrational mode register as $\ket{\phi_{i,0}}_i\rightarrow\ket{\phi_{i,1}}_i$. This can be implemented using e.g. generic state preparation or optimized variational circuits. This is done for $i=1,\cdots,M$.
    \item Execute the inverse operations in steps $4$ and $5$, effectively uncomputing those operations and returning the associated ancillas (registers $\ket{}_c$ and $\ket{}_I$) to their all-zeros computational basis states.
\end{enumerate}

All additional qubits required by this routine will be returned to the computational all-zeros basis state, which can then be re-used for other routines for the remainder of the computation. 

Despite the exponential cost of employing generic state preparation, the product state nature of the initial ground-state state makes the required circuits to only act on registers corresponding to a single vibrational mode, which consist of a small number of qubits. As such, even in the case where all the required operations are implemented with this approach, the cost of the initial state preparation is negligible when compared to that of implementing the time-evolution. The cost of preparing an initial state with $D$ independent product states using the approach above for a system with $M$ modes and $N$ modals per mode can be roughly estimated as follows. The cost of the first two uses of QROM in steps (4) and (5) together with the multicontrolled rotations of step (6) is roughly $D \log D$ \cite{ini_state} Toffoli gates (for simplicity, we multiply by 4 to convert to T-gates). To estimate the cost of step (3), we can consider the straightforward generic state preparation approach \cite{mottonen}: while the cost is exponential -- roughly $2^{N}$ single-qubit rotations are required -- the number of qubits is small in practice and only logarithmic in the maximum achievable modal cutoff $N_{\text{max}}$. The number of T-gates to implement a single-qubit rotation will vary depending on the decomposition and precision requirements: we can take one rotation to be implemented using 100 T-gates which corresponds to an accuracy of $\sim10^{-10}$ per rotation \cite{t_count}. Putting all this together, the full cost of state preparation is $100 \cdot 2^{N} M + 4\cdot D \log(D)$. In a typical system of interest in the current manuscript such as azidoacetylene, $M = 12$ and $D$ can be taken conservatively to be 1,000: for $N$, a typical value for modal cutoff is $5-10$, which does not scale with system size. Plugging these number into the expression, we find the cost to be $\sim 10^{4-6}$. Considering a total of $\sim 10^5$ circuit repetitions, this entails an additional cost of $\sim 10^{11}$ T-gates, which is dwarfed by the cost of time-evolutions of the same molecule corresponding to $\sim 10^{13}$ T-gates (cf. \cref{sec:benchmark}). 

Finally, we note that the initial state can often become more complex, such as in cases where a more sophisticated \textit{ansatz} than VSCF is required to capture the ground state, and/or as higher degrees of the Taylor expansion of the dipole operator are included. In these cases, the approach shown here can be extended to prepare such states.

\subsection{Caching technique for polynomial implementation} \label{app:caching}

Motivated by the fact that higher-order mode coordinate products occurring in the potential energy, for example $\hat q_i \hat q_j \hat q_k^2$, have sub-factors that are also present in the Hamiltonian, such as $\hat q_i \hat q_j$ and $\hat{q_k^2}$, and thus are also calculated in the process of performing time-evolution, we present here a concrete, heuristic scheme for \textit{caching} the results of the intermediate calculations such that they can be re-used rather than recalculated each time, saving the total number of multiplications to be performed (the number of additions will remain the same). For concreteness we limit ourselves to the case of a $2$-mode, fourth-order Taylor Hamiltonian, noting that the scheme can be straightforwardly extended to higher orders and couplings. We could use additional registers to cache more intermediate products and diminish the number of operations, however in this work we adopt a caching algorithm that will not use any additional qubits. We thus employ the main system register which consists of five ancilla registers of size $b_k = 10$ to store intermediate multiplication results, and a resource state of size $b_r = 25$ against which we perform the phase gradient technique \cite{phase_gradient}. In general, one needs $T+1$ ancilla registers of size $k$ for a Hamiltonian of Taylor order $T$. The particular scheme we use for resource estimation in this paper is as follows:
\begin{enumerate}
    \item Begin by computing a square $\hat q_m^2$ using the out-of-place squaring primitive and caching it in one of the ancilla registers. Usually all such terms are nonzero because of the harmonic part of the Hamiltonian.
    \begin{enumerate}
        \item Apply a multiply primitive against the cached result and the original system register to produce the corresponding $\hat q_m^3$ term, and cache the result in the second ancilla register.
        \begin{enumerate}
            \item Apply another multiply primitive to obtain $\hat q_m^4$ to place it in the third register, load the coefficient into the fourth register, then combine them in the fifth register and use the phase gradient trick against the resource state.  Finally, clean the third, fourth and fifth register.
            \item Using the same cached result of $\hat q_m^3$, implement all of the nonzero $\hat q_m^3 q_l$ terms in the Hamiltonian in the same manner as above, using the third, fourth and fifth register for the final product result, the coefficient, and the final full term, respectively. Clean the third, fourth and fifth register.
        \end{enumerate}
    \item Clean the second register.
    \item Now consider all of the nonzero terms of the form $\hat q_m^2 \hat q_l$ -- using the cached result of $\hat q_m^2$, compute and cache them in the second register.
        \begin{enumerate}
            \item Apply another multiply primitive to obtain $\hat q_m^2 q_l^2$ to place it in the third register, load the coefficient into the fourth register, then combine them in the fifth register and use the phase gradient trick against the resource state.  Finally, clean the third, fourth and fifth register.
            \item Using the same cached result of $\hat q_m^2 \hat q_l$, implement all of the nonzero $\hat q_m^2 \hat q_l \hat q_k$ terms in the Hamiltonian in the same manner as above, using the third, fourth and fifth register for the final product result, the coefficient, and the final full term, respectively. Clean the third, fourth and fifth register.
        \end{enumerate}
    \end{enumerate}    
    \item Finally, do all the required sequential multiplications to implement terms of the form $\hat q_m \hat q_l \hat q_k \hat q_n$.
\end{enumerate}
Given this scheme and the vibrational Hamiltonian, we directly count the required number of multiplications and additions needed to implement a single Trotter time-evolution step.

\section{Simulation details} \label{app:details}
In this appendix we discuss all the details for the NIR spectra simulations related to \cref{fig:h2s_full,fig:trotter,fig:hyper}. Note that the additional ancilla cost of several optimizations, such as the usage of quantum arithmetic circuits for implementing the exponentials in the Trotterization terms, makes the computational requirements of simulating these circuits significantly larger than what is possible via direct simulation of the wavefunction over all qubit registers. As such, a no-ancilla based approach was used for these simulations, in the sense that we directly encoded the exponentials of the kinetic and potential energy terms. However, given the precision guarantees of the quantum arithmetic and phase-gradient-based exponentiation routine, the only difference relative to our ``exact'' simulations should appear in the rounding of the coefficients that appear in the exponentials. The accuracy of this rounding will be dictated by the number of qubits used in the phase-gradient register and the associated ``addition register'' where the coefficient to be exponentiated is loaded. By using 25 qubits for the phase-gradient register, an accuracy of $3.0\times 10^{-8}$ is achieved for these operations, making the recovered spectra indistinguishable from the exact simulations performed classically with floating point arithmetic. Different polynomial orders appearing in the Taylor expansion of the Hamiltonian in \cref{eq:taylor_ham} typically have different orders of magnitude associated to their coefficients, e.g. $\Phi^{(2)}_{ij}$ and $\Phi^{(4)}_{ijkl}$, which from a first glance could appear as needing a high number of qubits $k$ in the associated coefficient register which loads these coefficients (see \cref{fig:poly_evolution}. However, we are using a different resource state to implement each of the terms coming from different orders. As such, the associated constant $\phi_b$ appearing in \cref{eq:resource} can be adjusted to target the specific scale of the polynomial order it is implementing. This translates into a smaller accuracy requirement for this qubit register, which will then yield the relative resolution with which coefficients of the same polynomial order are being implemented. \\

Vibrational Hamiltonians were generated using the PennyLane tools for vibrational system generation \cite{pennylane}. Internally, this routine calculates the PES using the restricted Hartree-Fock method with the 6-31G basis via the PySCF package \cite{pyscf_1,pyscf_2}. First the vibrational normal modes are determined at the equilibrium geometry by standard harmonic analysis. Next, mode localization is performed as described in \cref{app:mode_loc}, obtaining a new set of vibrational modes. A Gauss-Hermite quadrature-based grid with 9 points is associated to each vibrational mode, generating the PES by performing single-point ground-state energy calculations at system geometries set by sweeping through all the allowable combinations of those grid points across the modes. Since we focused on the two-mode expansion, this means in each geometry at most two vibrational modes were offset from equilibrium simultaneously. As shown in Ref.~\cite{mode_loc_1} and discussed in \cref{app:mode_loc}, performing a $2$-mode expansion using localized modes yields accuracy from a higher order expansion using normal modes, while greatly reducing the number of terms in the Hamiltonian. The coefficients for the Hamiltonian in \cref{eq:taylor_ham} are then obtained through a linear regression fit of the obtained PES going up to a fourth-order Taylor expansion. The $2$-mode, fourth-order Taylor Hamiltonian is widely used in vibrational spectroscopy calculations and is often referred to as the quartic force field approach (QFF). However, it can also be advantageous to use sixth or higher orders for increasing the quality of the obtained spectra. The quantum-arithmetic-based implementation of the time-evolution alongside the caching technique discussed in \cref{app:caching} allows for an efficient implementation of higher order Taylor expansions, as discussed in \cref{app:scaling}. We also note that using more accurate electronic structure methods has been shown to be crucial for recovering accurate vibrational spectra \cite{pes_quality}. However, for the purposes of this work we expect the Hamiltonians obtained using the lower level of theory to be representative of those that would be obtained with more accurate methods. The resource estimates obtained in this work should thus be extremely similar to ones for Hamiltonians from higher levels of electronic structure. 
    
\end{document}

%% file: preamble.tex
\usepackage{amsthm}
\usepackage{amsmath}
\usepackage{amsfonts}
\usepackage{mathtools}
\usepackage{xcolor}
\usepackage{graphicx}
\graphicspath{{figures/}}
\usepackage[T1]{fontenc}
\usepackage{centernot}
\usepackage{nicefrac}
\usepackage{float}

\newcommand{\bra}[1]{\langle #1 \rvert}
\newcommand{\ket}[1]{\lvert #1 \rangle}

\newcommand{\mhat}[1]{\hat{\mathcal{#1}}}

 \usepackage{tikz}
 \usetikzlibrary{quantikz2}

\usepackage{glossaries}
\newacronym{QPE}{QPE}{quantum phase estimation}
\newacronym{GQPE}{GQPE}{generalized quantum phase estimation}
\newacronym{LCU}{LCU}{linear combination of unitaries}
\newacronym{BCH}{BCH}{Baker-Campbell-Hausdorff}
\newacronym{QFT}{QFT}{quantum Fourier transform}
\newacronym{PES}{PES}{potential energy surface}
\newacronym{NIR}{NIR}{near-infrared}
\newacronym{VSCF}{VSCF}{vibrational self-consistent field}

\DeclareMathOperator*{\argmax}{argmax}